\documentclass[useAMS,usenatbib]{mn2e}
\usepackage{multirow}
\usepackage{amsmath}
\usepackage{amssymb}
\usepackage{graphicx}
\usepackage{multirow}
\usepackage{subfigure}
\usepackage[usenames,dvipsnames]{xcolor}
\newcommand{\appropto}{\mathrel{\vcenter{
  \offinterlineskip\halign{\hfil$##$\cr
    \propto\cr\noalign{\kern2pt}\sim\cr\noalign{\kern-2pt}}}}}
\begin{document}

\title{Galaxy gas ejection in radio galaxies: the case of 3C\,35}

\author[E. Mannering et al.]
{E.~Mannering$^{1}$\thanks{l.mannering@bristol.ac.uk}, D. M.~Worrall$^{1}$, M.~Birkinshaw$^{1}$ \\
$^{1}$H.H. Wills Physics Laboratory, University of Bristol, Tyndall Avenue,  Bristol BS8 1TL, UK } 

\maketitle

\label{firstpage}

\begin{abstract}
We report results from {\em XMM-Newton} and {\em Chandra}  observations of the nearby ($z=0.067$) giant radio galaxy 3C\,35. We find evidence for an X-ray emitting gas belt, orthogonal to and lying between the lobes of 3C\,35, which we interpret as fossil-group gas driven outwards by the expanding radio lobes. We also detect weak emission from a second, more extended group-type environment, as well as inverse-Compton X-ray emission from the radio lobes. The morphological structure of the radio lobes and gas belt point to co-evolution. Furthermore, the radio source is powerful enough to eject galaxy-scale gas out to distances of $100$\,kpc, and the ages of the two features are comparable ($t_{\rm{synch}}\approx140\,$Myr, $t_{\rm{belt}}\approx80$\,Myr). The destruction of 3C\,35's atmosphere may offer clues as to how fossil systems are regulated: radio galaxies need to be of power comparable to 3C\,35 to displace and regulate fossil-group gas. We discuss the implications of the gas belt in 3C\,35 in terms of AGN fuelling and feedback.
\end{abstract}

\section{Introduction}
\label{sec:introxray}
Radio galaxies are important sources of heating in groups and clusters, and feedback by active galactic nuclei (AGN) is now considered as the most likely mechanism to balance radiative cooling \citep[e.g.,][]{mcnamara2007}. The hot intergalactic medium (IGM) may be directly heated by shocks and sound waves driven by the expansion of radio jets \citep[e.g.,][]{kraft2012,forman2007}. Cavities inflated by the radio jets may also limit the ability of the IGM to cool, by doing work on the gas, lifting it out from the densest regions of the group or cluster \citep[e.g.,][]{blanton2009, birzan2008}. To understand the role of fuelling/feedback of AGN, it is necessary to study radio jet interactions on all scales, from AGN to group/cluster environments. Nearby radio galaxies are obviously best suited to this task.

One such source is 3C\,35, a low-redshift \citep[$z=0.067$,][]{Spinrad85}, low-excitation \citep{Jackson97} giant radio galaxy with an elliptical host. \citet[][hereafter O10]{Orru2010} discuss its \citet[][herein FR]{Fanaroff74} class II radio morphology. \citet{owen1989} measure a 178\,MHz radio power of $P_{178\,\mathrm{MHz}}=10^{24.99}$\,W\,Hz$^{-1}$\,sr$^{-1}$, spanning a total angular size of roughly $12.5'$\footnote{Corresponding to a linear size of 950\,kpc at $z=0.067$}. 
\citet{burns1977} classify 3C\,35 as lying in an open cluster, coincident to within 2 cluster radii of a Zwicky cluster \citep{Zwicky1961-68}, however \citet{mchardy1974} find this to be a weak cluster environment and suggest 3C\,35 probably resides in a small group. In contrast, \citet{guindon1979} classified 3C\,35 as a non-cluster source based on the definition of an Abell or Zwicky cluster.

In this paper, we report X-ray measurements of a gas belt that we argue may be a disrupted fossil-group X-ray halo, as well as diffuse emission in the lobes of 3C\,35. Constraining the spectral properties and energetics of the constituent relativistic particles in the lobes \citep[e.g.][for a review]{worrall2009} and the thermodynamical properties of the IGM provide useful clues as to the origin of the gas belt. 3C\,35 seems to be the first case where we see the X-ray halo of a fossil group being ejected following an interaction with the radio structure. 

\subsection{Fossil groups and radio galaxies}
Fossil groups are a class of system that has an extended, hot gaseous halo encompassing a single giant elliptical galaxy, but with the gravitating mass of a galaxy group \citep{hess2012, jones2003}. They are believed to be the merger remnants of a galaxy group whose $L_{*}$ galaxies have coalesced, leaving behind an extended gas halo and dwarf galaxies \citep[][and references therein]{dariush2007}. Fossil groups are identified observationally by two criteria; an extended X-ray source with $L_{\rm{X}}>10^{42}h^{-2}$\,erg\,s$^{-1}$ \citep{jones2003}, and an optically dominant elliptical galaxy, where the companion galaxies are fainter by two magnitudes in $R$. The lower limit in X-ray luminosity attempts to exclude normal bright elliptical galaxies exhibiting a hot coronal gas component, and the magnitude gap in $R$ ensures that a single elliptical dominates the system. 

$N-$body simulations suggest that fossil groups formed early \citep{donghia2005}, assembling half of their total dark matter mass at $z\gtrsim1$, with later growth by minor mergers only, whereas non-fossil groups formed much later. The extensive, virialised X-ray halo, lack of $\sim L_{*}$ galaxies and the large magnitude gap tend to support this interpretation \citep{jones2003, mendesdeoliveira2006}. Fossil groups are at least as numerous as all poor and rich clusters combined \citep{jones2003}, contributing significantly to the total mass density of the Universe. \citet{dariush2007} use the Millennium simulation and semi-analytic galaxy catalogues to trace the formation and evolution of fossil groups, and find that fossil groups may be a phase of hierarchical evolution, rather than a final stage of mass assembly \citep{labarbera2009}. 

Fossil-group halos are hotter and more X-ray luminous than non-fossil groups and clusters of similar gravitational mass \citep{khosroshahi2007}. \citet{croston2005} showed that radio-loud ($L_{\rm{1.4\,GHz}}>10^{23}\,$W\,Hz$^{-1}$) non-fossil groups are hotter at a given X-ray luminosity than radio-quiet groups, indicating AGN activity is heating the IGM. \citet{hess2012} found a weak correlation between 1.4\,GHz luminosity of radio-loud AGN and the X-ray luminosity of the halo, tentatively suggesting that the AGN contributes to the energy deposition in the IGM of fossil systems too. Two-thirds of their sample have radio-loud AGN, suggesting that AGN fuelling continues (or is re-ignited) long after the last major merger.

\subsection{Inverse Compton emission}
Lobe radio emission is well described by the synchrotron process (O10 for 3C\,35), while the extended non-thermal X-ray radiation from the lobes of radio galaxies and quasars detected in many FR II sources is attributed to the inverse-Compton (iC) mechanism \citep[e.g.,][]{Brunetti97,Hardcastle02,Comastri03,croston2005,Hardcastle05,Isobe05, Goodger08,Isobe11}. Modelling lobe-related iC X-ray emission enables constraints to be placed on the electron energy density ($u_{e}$) and magnetic field strength ($B$) which otherwise cannot be decoupled using radio data alone. X-rays from the lobes are a sensitive probe of $u_{e}$ in low-energy electrons within the lobes and, in combination with radio mapping, sometimes allow investigation of positional variations in $B$ and the electron energy spectrum \citep[e.g.,][]{croston2005, Goodger08}. 

Giant radio galaxies enable exploration of the late phase in the evolution of radio jets, as lobe energetics are an important indicator of past activity because they accumulate the energy deposited by the jets. If the non-thermal particle and magnetic field energy densities can be measured from the diffuse (iC) lobe emission, estimates can be made of the total energy (which is potentially available to be transferred to the environment) and the distribution of internal energy within the source, without assuming equipartition. This in turn constrains the past duration of the active phase of the active nucleus. 

\subsection{3C\,35}
3C\,35 is an old source: its spectral age is estimated by O10 as $143\pm20$\,Myr. It has been previously observed with the X-ray Imaging Spectrometer (XIS) on board \textit{Suzaku} \citep{Mitsuda07}, revealing faint extended X-ray emission associated with its radio lobes.  \citet[][hereafter IS11]{Isobe11} integrated the \textit{Suzaku} spectrum within a large rectangular region containing the whole radio structure of 3C\,35. They report iC emission associated with the lobes (power-law index $\Gamma=1.35^{+0.56  +0.11}_{-0.86  -0.10}$, where the first errors are due to signal statistics, and the second are from background uncertainties), as well as a soft thermal component corresponding to thermal plasma emission from the host galaxy ($kT=1.36^{+0.28  +0.02}_{-0.23  -0.02}$\,keV). In the case of giant radio galaxies such as 3C\,35, the cosmic microwave background (CMB) is the dominant source of photons to be iC scattered (by relativistic electrons of $\gamma\sim10^{3}$) up to X-ray energies, although nuclear photons and starlight are important in some sources where they are scattered by lower-energy electrons \citep[e.g.,][]{Brunetti97, stawarz2003}. Due to low electron densities in the large lobe volumes, the synchrotron self Compton process makes a negligible contribution to the X-ray flux \citep[e.g.,][]{croston2005, Hardcastle02}. 

We use new \textit{XMM-Newton} (hereafter {\em XMM}) observations and archival \textit{Chandra} data to investigate an X-ray gas belt seen orthogonal to the lobes of 3C\,35, and to probe the iC properties of the lobes themselves. In Section \ref{sec:obsdata} we briefly describe the \textit{XMM} and \textit{Chandra}  observations and data reduction. The X-ray morphology of 3C\,35, and the results of spectral fitting are discussed in Sections \ref{sec:morphology} and \ref{sec:spectroscopy_results}. In Section \ref{sec:discussion} we present a discussion of our findings. Throughout, we use a flat $\Lambda$CDM cosmology with $\Omega_{m_{0}}=0.3$ and $\Omega_{\Lambda 0}=0.7$. We adopt $H_{0}=70$ km\,s$^{-1}$\,Mpc$^{-1}$.

\section{Observations and Data Reduction}
\label{sec:obsdata}
We used archival NRAO Very Large Array (VLA) and {\em Chandra} data in combination with new
{\em XMM} X-ray data to examine the emission mechanisms in the lobes of 3C\,35 and probe the physical parameters of the gas belt. Our X-ray data
have improved sensitivity and resolution compared to the
{\em Suzaku} data of IS11. The high spatial resolution of \textit{Chandra} enables us to reduce contamination from X-ray background sources and to isolate individual source components. However, the small {\em Chandra} effective area leads to a low number of ACIS counts, whereas {\em XMM} count rates are higher, enabling analysis of the properties of the X-ray emission and electron population as a function of position. 
\subsection{XMM-Newton}
\label{subsec:obsxmm}
\subsubsection{Data reduction}
We observed 3C\,35 with {\it XMM} for just under $100$\,ks during February 2011. Data are presented from the EPIC MOS1, MOS2 and pn cameras in full-frame mode using the {\sc medium} filter.
We reprocessed the Observation Data Format (ODF) files\footnote{Data were downloaded from the {\it XMM} Science Operations Centre (SOC) and analysed using SAS version 11.0.0} using the dedicated pipeline chain tasks {\fontfamily{pcr}\selectfont emchain} and {\fontfamily{pcr}\selectfont epchain} for EPIC MOS and pn, respectively. We extracted single and double events ({\sc pattern} 0-4) for the pn data and up to and including quadruple events ({\sc pattern} 0-12) for the MOS data. We make use of the \textit{XMM} Extended Source Analysis Software procedure \citep[XMM-ESAS;][]{Snowden2011} and the associated current calibration files (CCF) which contain filter wheel closed (FWC), quiescent particle background (QPB) and soft proton (SP) calibration data\footnote{ftp://xmm.esac.esa.int/pub/ccf/constituents/extras/esas\_caldb/}.

The SP component due to solar flares cannot be well removed via background subtraction. It is highly variable, causing the spectrum to change rapidly with time. Frames with high count rates must be excluded, which unfortunately can significantly reduce net exposure time. 
\subsubsection{Flare cleaning}
\begin{figure}
\includegraphics[width=\columnwidth]{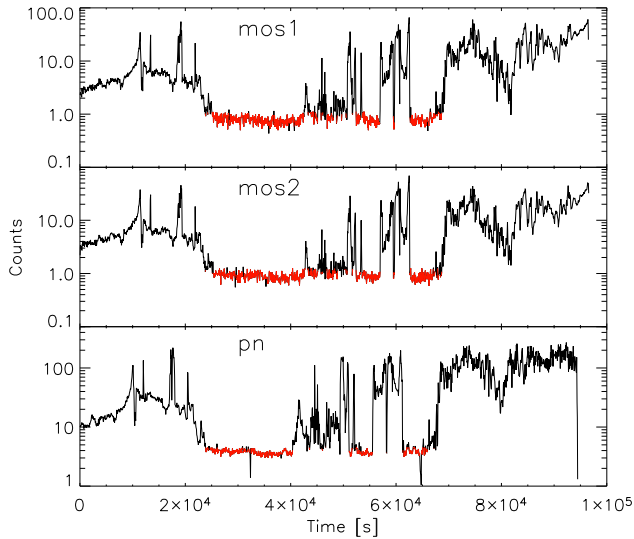}
\caption{Light curves generated from the 3C\,35 MOS1, MOS2 and pn event lists. The black line shows the raw data in 60\,s bins for the whole field of view. The red lines show the GTI applied `clean' data. }
\label{fig:lightcurvesxmm}
\end{figure}

In order to create a Good Time Intervals (GTI) file and ascertain the severity of the increased solar activity on the observation of 3C\,35, we followed the prescription of background screening outlined in the XMM-ESAS guide \citep{Snowden2011} and ran the dedicated ESAS tasks {\fontfamily{pcr}\selectfont mos-filter} and {\fontfamily{pcr}\selectfont pn-filter} to generate SP contamination-filtered products for the field-of-view data. It is clear from the light curves (Figure \ref{fig:lightcurvesxmm}) that high flaring dominated much of the observation. A Gaussian was fitted to the nominal count rate distribution and a GTI was created for those time intervals with count rates within a threshold value of an acceptable level ($\pm1.5\sigma$, Table \ref{table:3sigclip}).
\begin{table}
\begin{center}
\caption{{\em XMM} exposures. The duration refers the total livetime of the observation. Net exposure is the good time remaining after $1.5\sigma$ background screening.}
\begin{tabular}{@{}lcccc}
\hline
Detector & Duration (ks) & Net exposure (ks) \\
\hline
MOS1 & 95.4 &30.7\\
MOS2 & 95.4 &31.3\\
pn & 79.4 & 21.4\\
\hline 
\end{tabular}
\label{table:3sigclip}
\end{center}
\end{table}
CCD 4 on the MOS1 detector was operating in an anomalous state (the background at E $<1$\,keV is strongly enhanced) and was excluded from subsequent processing, along with CCD 6, which previously suffered micro-meteorite damage. The XMM-ESAS task {\fontfamily{pcr}\selectfont cheese} combines MOS and pn data to generate point-source lists (down to a limiting flux of $1.0\times10^{-15}$\,erg\,cm$^{-2}$\,s$^{-1}$), exposure maps and mask images for use in creating source-excluded spectra. Tasks {\fontfamily{pcr}\selectfont mos-spectra} and {\fontfamily{pcr}\selectfont pn-spectra} were run in order to extract spectra from the cleaned event files for a given selection region. Finally, the tasks {\fontfamily{pcr}\selectfont mos\_back} and {\fontfamily{pcr}\selectfont pn\_back} were run to create model particle background spectra for the selected regions.
The \textit{XMM} background was estimated using methods described in \S{\ref{subsubsec:xmm_back}}. 
Spectra were binned using a minimum of 50 counts per bin for the MOS and pn instruments, over the energy range 0.4-7.2\,keV. 
\subsection{Chandra}
\label{subsec:obschandra}
3C\,35 was observed with {\it Chandra} using the Advanced CCD Imaging Spectrometer (ACIS) on 2009 March 8. The front-illuminated ACIS-I3 CCD was placed at the focus of the observation, configured in timed exposure and VFAINT telemetry mode. 
The data were reduced using the latest \textit{Chandra} software ({\sc ciao} 4.4 and CALDB 4.4.10) following the \textit{Chandra} analysis threads\footnote{http://cxc.harvard.edu/ciao/threads/index.html}. 
We reprocessed the level 1 event files, applying {\sc vfaint} filtering. The latest charge-transfer inefficiency (CTI) correction and time-dependent gain adjustment were applied and a background light curve was produced. The observation did not suffer from background flaring, and we selected events of grade 0, 2, 3, 4 or 6 resulting in a net exposure time of 25.63\,ks. The {\sc ciao} function {\sc wavdetect} was used to identify point sources from a 0.3-12.5\,keV exposure corrected image, and regions around 62 sources over 4 CCDs were excluded from subsequent spectral extractions. X-ray spectra were extracted from regions using the {\sc specextract} task, and grouped to a minimum of 30 counts per bin. 
\subsection{VLA}
VLA data taken as part of program AL419 were downloaded from the archive. We use 1.42\,GHz data from an observation made on 1997 August 27th in C array. The radio map was made using standard AIPS tasks, with phase self calibration and a restoring beam of $14\times14$\arcsec (FWHM). Data were combined using roughly 3\,hr of on-source integration time to obtain a root-mean-square (rms) sensitivity $<0.07$\,mJy/beam across the field. Contours are shown in Figure \ref{fig:lobes}. 

\section{X-ray Morphology}
\label{sec:morphology}
\begin{figure}
\begin{center}
\includegraphics[width=3.0in]{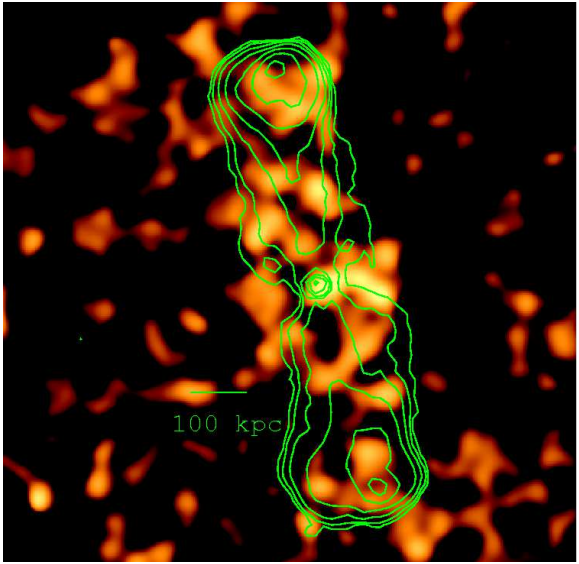}
\caption{\textit{Chandra} exposure-corrected image of 3C\,35 in the energy range 0.5-7.0\,keV. VLA C-array radio contours at 1.42\,GHz are overlaid. The peak radio flux is 32.0\,mJy/beam and contour levels are spaced by a factor of 2, the lowest being 0.5\,mJy/beam (restoring beam size $\approx14\times14$\arcsec\,\,FWHM). The X-ray data are binned by a factor of 4 into pixels of size roughly $2\arcsec$. Point sources (including the X-ray nucleus) were removed and the values interpolated from the surrounding pixels using the {\sc ciao} tasks {\sc wavdetect} and {\sc dmfilth}. Data are smoothed using the {\sc ciao} task {\sc aconvolve} with a two dimensional Gaussian function of $\sigma=10$ pixels ($\approx20\arcsec$).}
\label{fig:lobes}
\end{center}
\end{figure} 
A heavily smoothed 0.5-7.0\,keV, background-inclusive \textit{Chandra} image with point sources excluded is shown in Figure \ref{fig:lobes}, with radio contours obtained from our VLA map at 1.4\,GHz overlaid. The X-ray core is weak (we detected $9$ counts in the {\em Chandra} data) and offset from the radio nucleus (O10) by approximately $0.06$\arcsec, which is within the systematic uncertainty of \textit{Chandra} X-ray astrometry. In order to study the morphology of the extended emission, we excluded all point sources in the field, including the core. {\sc dmfilth} was used to fill in the gap regions with intensities from surrounding areas. Figure \ref{fig:lobes} shows the result after smoothing using the CIAO tool {\sc aconvolve}, with a Gaussian function of $\sigma\approx20\arcsec$ as a convolution kernel, and dividing by the exposure map in order to correct for varying effective area across the field. A small X-ray source is offset from the radio core; this is extended emission, the X-ray nucleus was properly subtracted within an exclusion region of radius 3.5\arcsec. 

\begin{figure}
\begin{center}
\includegraphics[width=1.5in, height=2.6in]{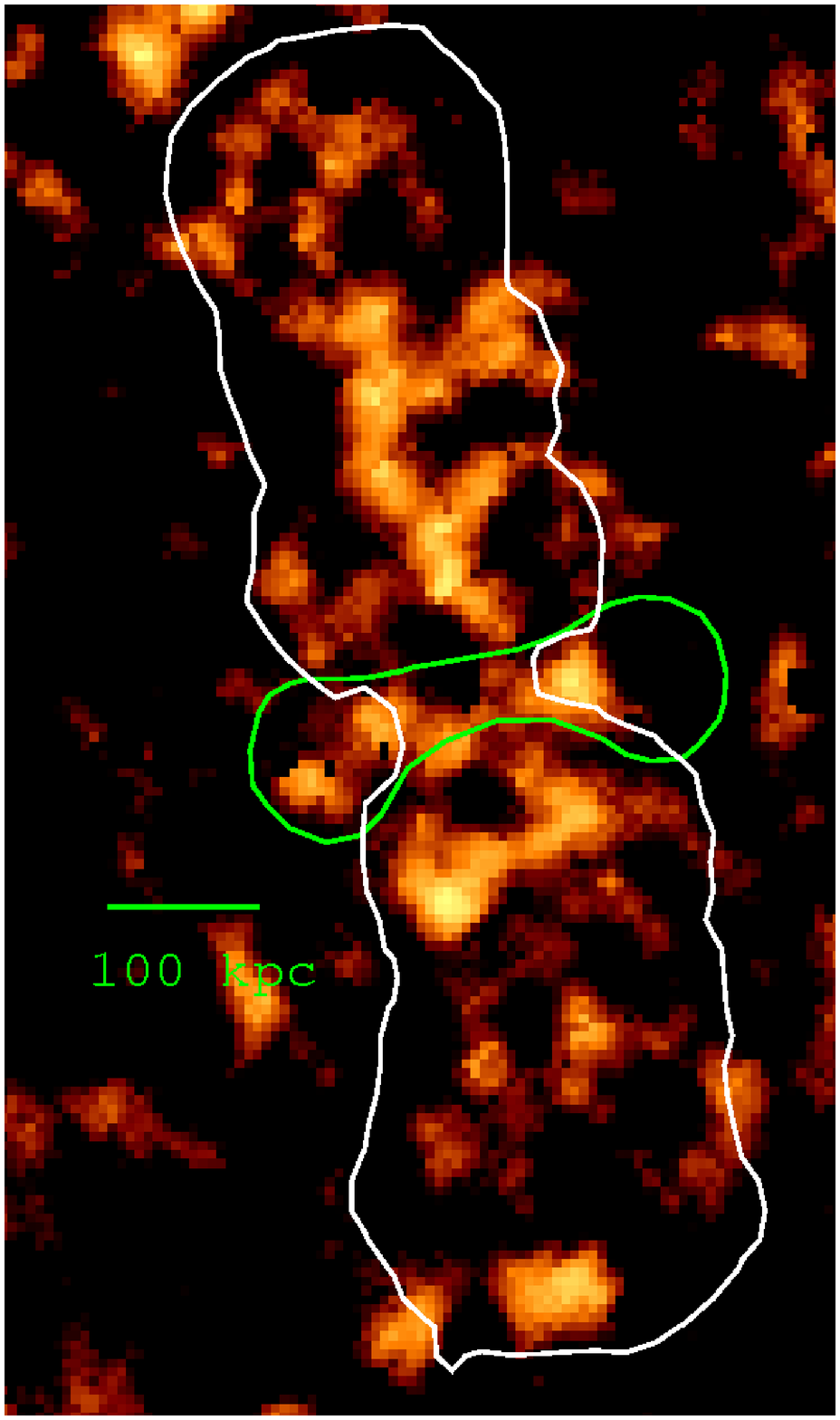}
\includegraphics[width=1.5in, height=2.6in]{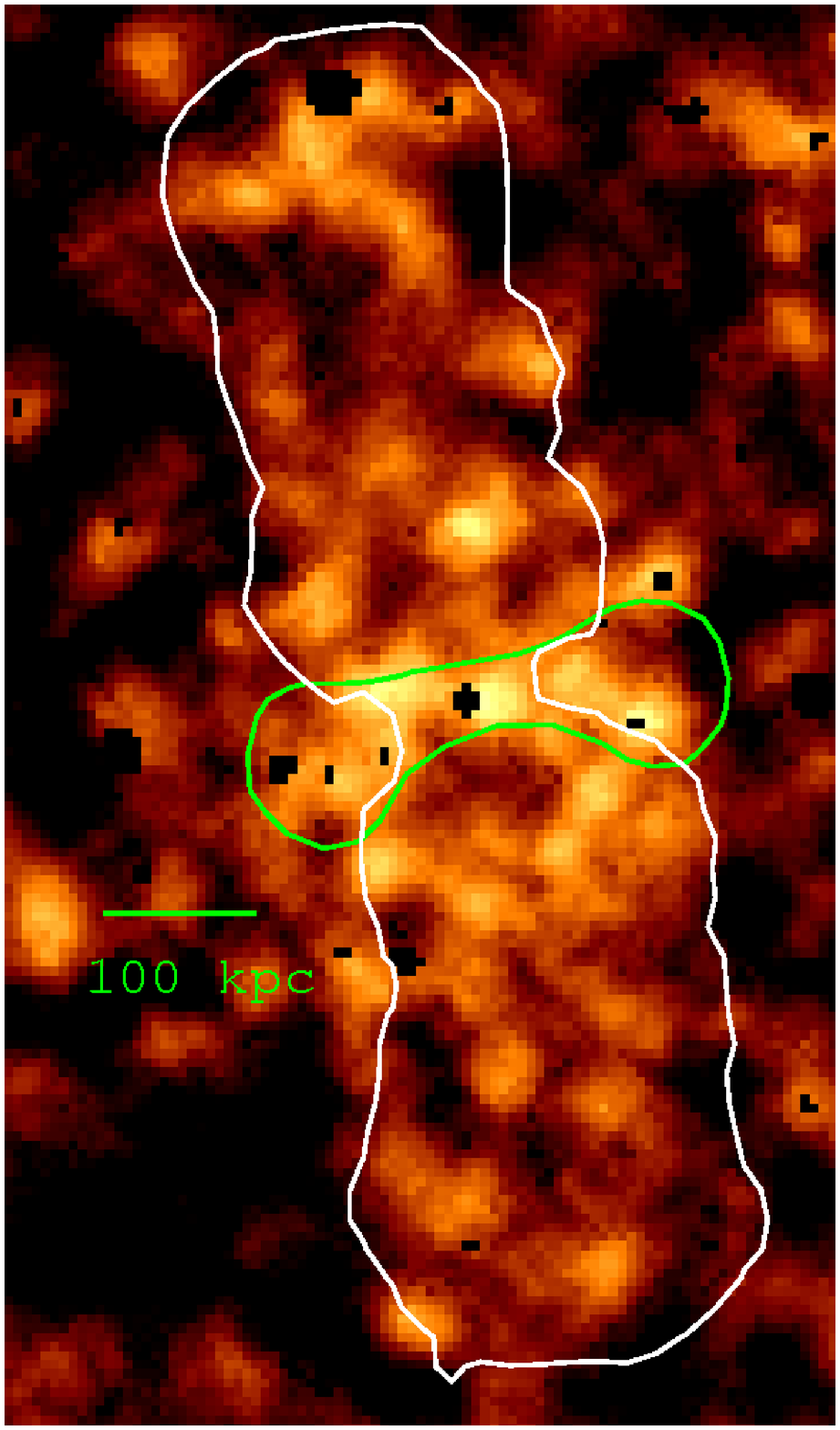}
\caption{Exposure corrected and background subtracted MOS+pn \textit{XMM} 2.0-7.0\,keV {\em (left)} and 0.4-1.25\,keV {\em (right)} images. Point sources in the field have been removed, the data are binned into 5\arcsec pixels and adaptively smoothed by a kernel of 50 counts. Overlaid are the lowest radio contour from Figure \ref{fig:lobes} and the region used to define the gas belt.}
\label{fig:imgxmm}
\end{center}
\end{figure}

We created an {\em XMM} image in the soft energy band $0.4-1.25\,$keV and the hard band $2.0-7.0$\,keV (the emission is dominated by instrumental lines at intermediate energies). We applied a correction for residual SP contamination to the data and combined the images and background files from all three detectors. 
Figure \ref{fig:imgxmm} shows the  exposure corrected MOS+pn image, binned into 5\arcsec pixels and adaptively smoothed. We used an off-source region to estimate an average surface brightness of the X-ray background (described in \S\ref{subsubsec:xmm_back}), and subtracted this value from the image. 

In both the {\em Chandra} and {\em XMM} data there is a clear detection of faint, diffuse X-ray emission filling the radio lobes, suggesting a non-thermal inverse-Compton origin as previously reported by IS11 on the basis of lower-resolution \textit{Suzaku} data. There is no excess X-ray emission associated with the radio hotspots. The striking filamentary appearance of the X-ray emission in Figure \ref{fig:imgxmm} (left) may be an artefact of the smoothing, but is evocative of a jet-like channel extending to the edge of the northern hot-spot. With the low number of counts, however, we cannot pursue this further at present. 

\subsection{Point-spread function}
\begin{figure}
\begin{center}
\includegraphics[width=2.7in]{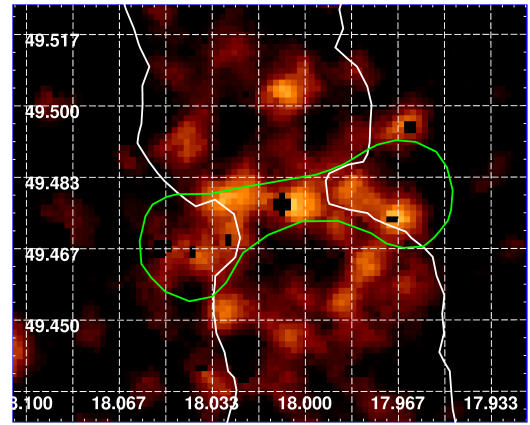}
\caption{{\em XMM} exposure corrected and background subtracted MOS+pn 0.4-1.25\,keV image of the gas belt. The data are binned to 5\arcsec and adaptively smoothed by a kernel of 50 counts, after point sources have been removed. Gridlines are spaced at $\Delta RA, \Delta \delta=(1.0\arcmin, 1.0\arcmin)$ and are marked in degrees. }
\label{fig:xmmgaszoom}
\end{center}
\end{figure}
The {\em XMM} PSF is a strong function of off-axis angle and photon energy \citep{lloyddavies2011}, becoming asymmetric at very large off-axis angles. It is important to know the variation of the PSF across the FOV in order to correctly define extraction areas for point sources in the gas belt. The belt extends roughly $2\arcmin$ from the centre of the field of view. In table 2 of \citet{xmmhandbook2_9}, the on-axis in orbit 1.5\,keV PSFs are given as having a FWHM $\lesssim12.5$\arcsec for the pn telescope and $\lesssim4.4$\arcsec for the MOS1 and 2 detectors. 
In figure 9 of \citet{xmmhandbook1_1} the radius at which 90$\%$ of the total energy of a point-source is encircled (W90) is plotted against off-axis angle for different energies. W90 for 1.5\,keV at $2\arcmin$ is approximately 15\arcsec. 

We modified the {\em XMM} masks to include also all sources detected in the {\em Chandra} data, increasing the minimum radius of a point-source to 15\arcsec\, in order to take into account the PSF increase off-axis. Less than half of the point sources detected in the {\em Chandra} data were detected in the {\em XMM} data. Figure \ref{fig:xmmgaszoom} shows the gas belt in the {\em XMM} data after point sources have been excised and the data have been heavily smoothed. There is a clear detection of X-ray counts in the belt region. 

\subsection{Gas belt}
\label{subsec:gasbelt_dss}
In both the {\em XMM} and {\em Chandra} data, particularly at soft energies, we find evidence of an X-ray gas belt constraining the central radio emission from the lobes of 3C\,35. Gas cavities associated with the radio lobes are not seen, in contrast to many cluster sources \citep[e.g.,][]{Rafferty2006}, and no extended thermal emission has been previously separated spatially in the field of 3C\,35. The belt emission is detected out to about 170\,kpc from the core, and it is coincident with a deficit in the radio emission (Figures \ref{fig:lobes} and \ref{fig:imgxmm}). We defined a region for spectral extraction by eye, making use of the radio contours as a guide to capture the `dog-bone' shape of the belt emission. 
\begin{figure}
\begin{center}
\includegraphics[width=1.2in]{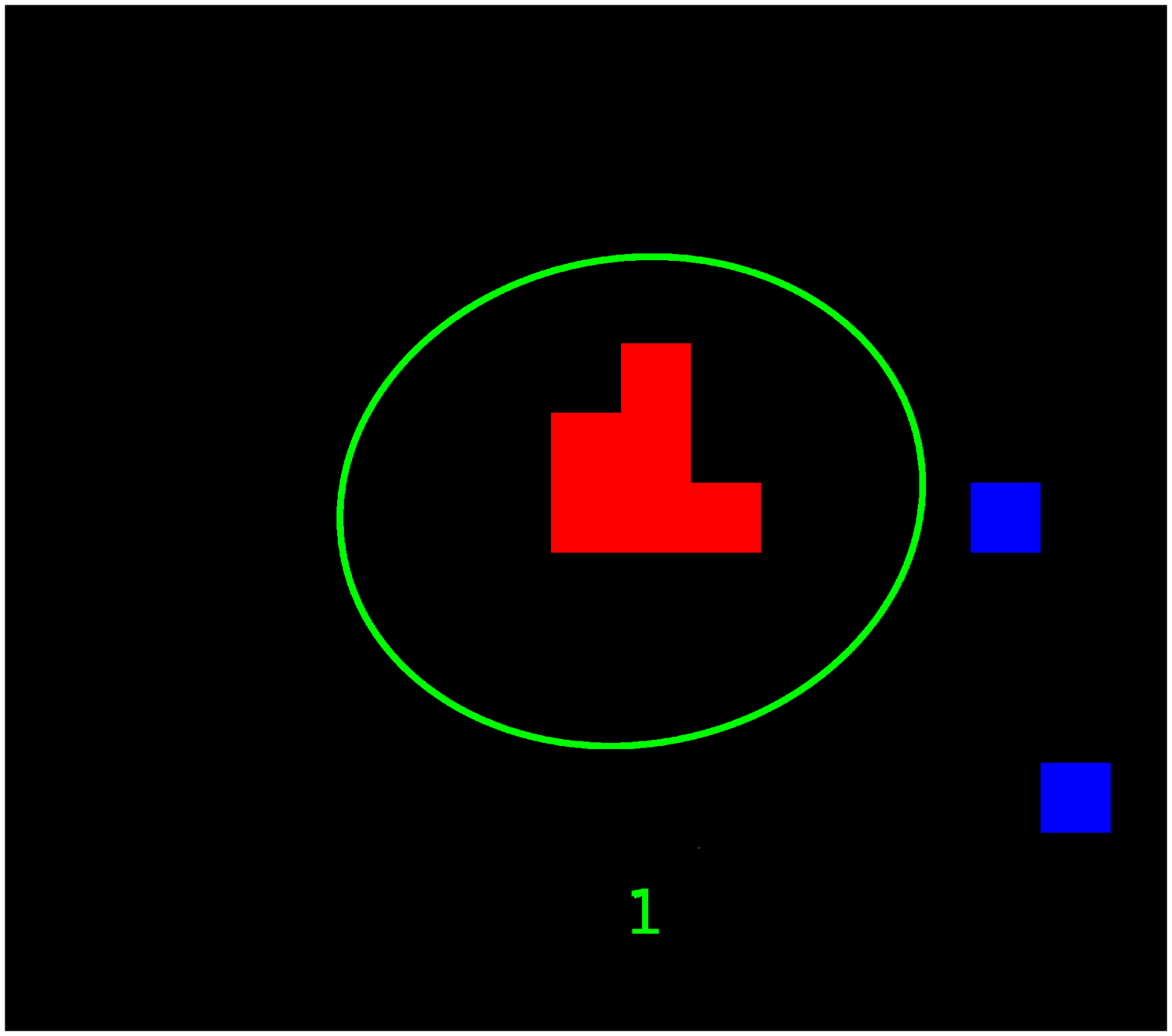}
\includegraphics[width=1.2in]{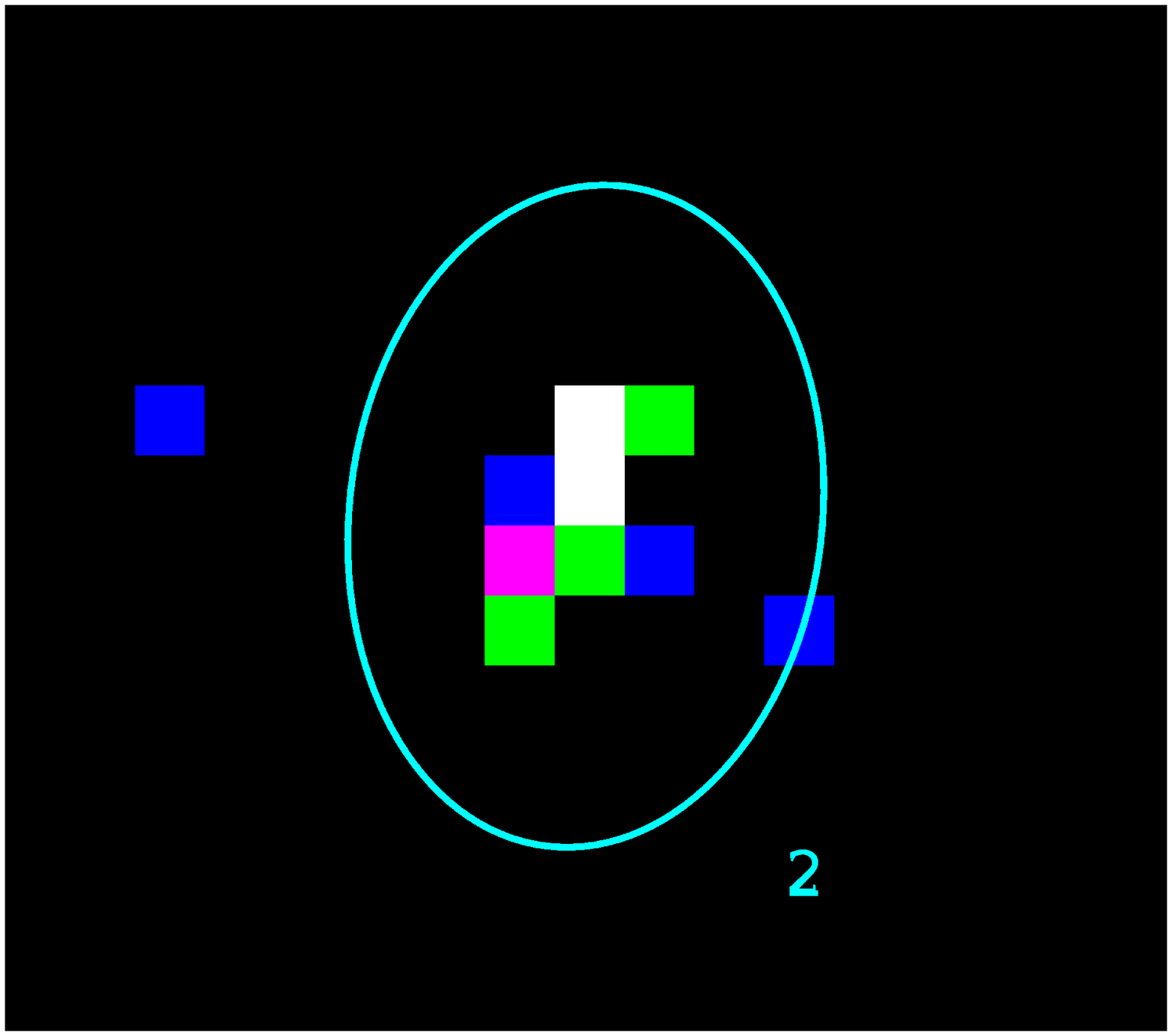}
\includegraphics[width=1.2in]{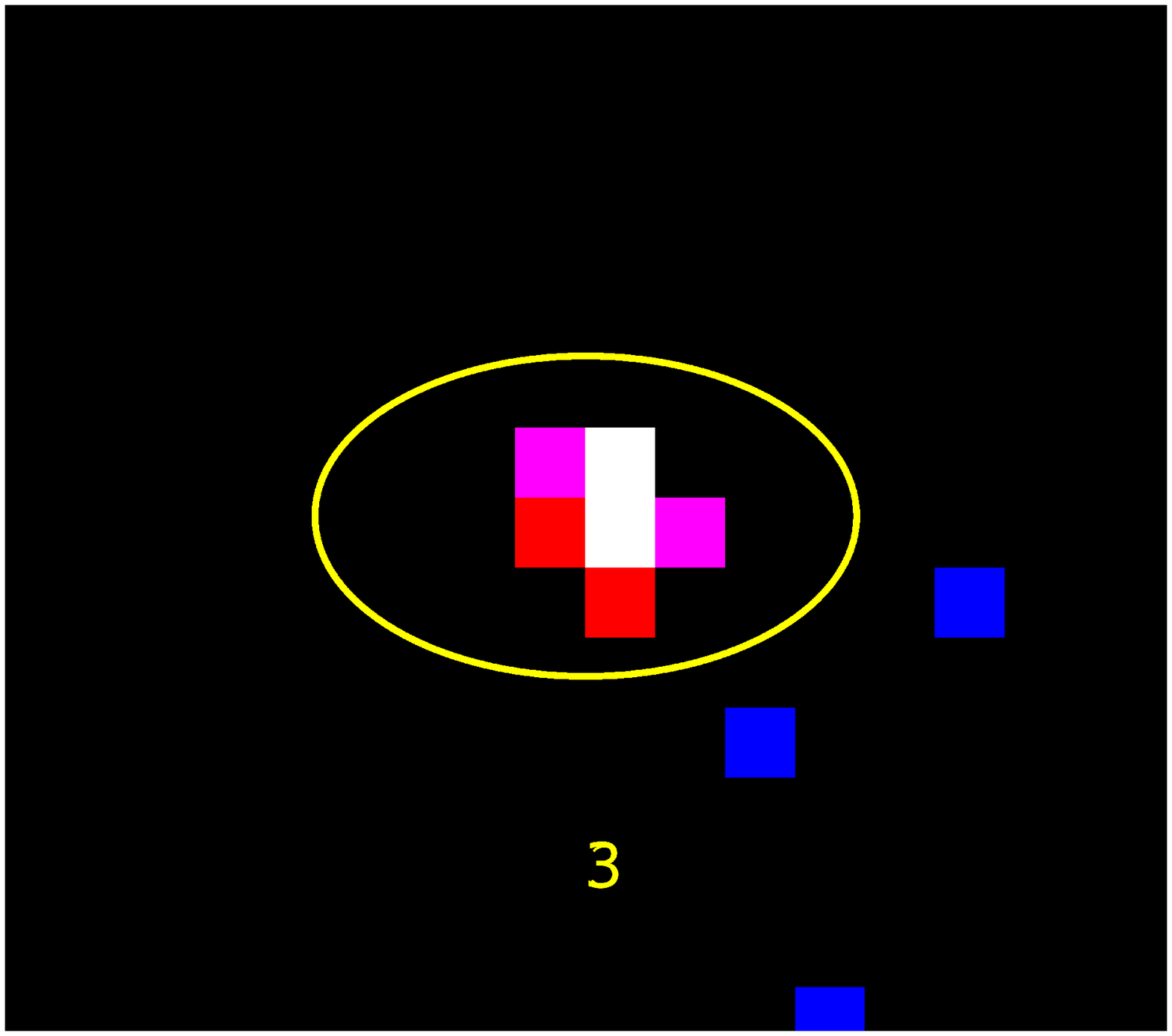}
\includegraphics[width=1.2in]{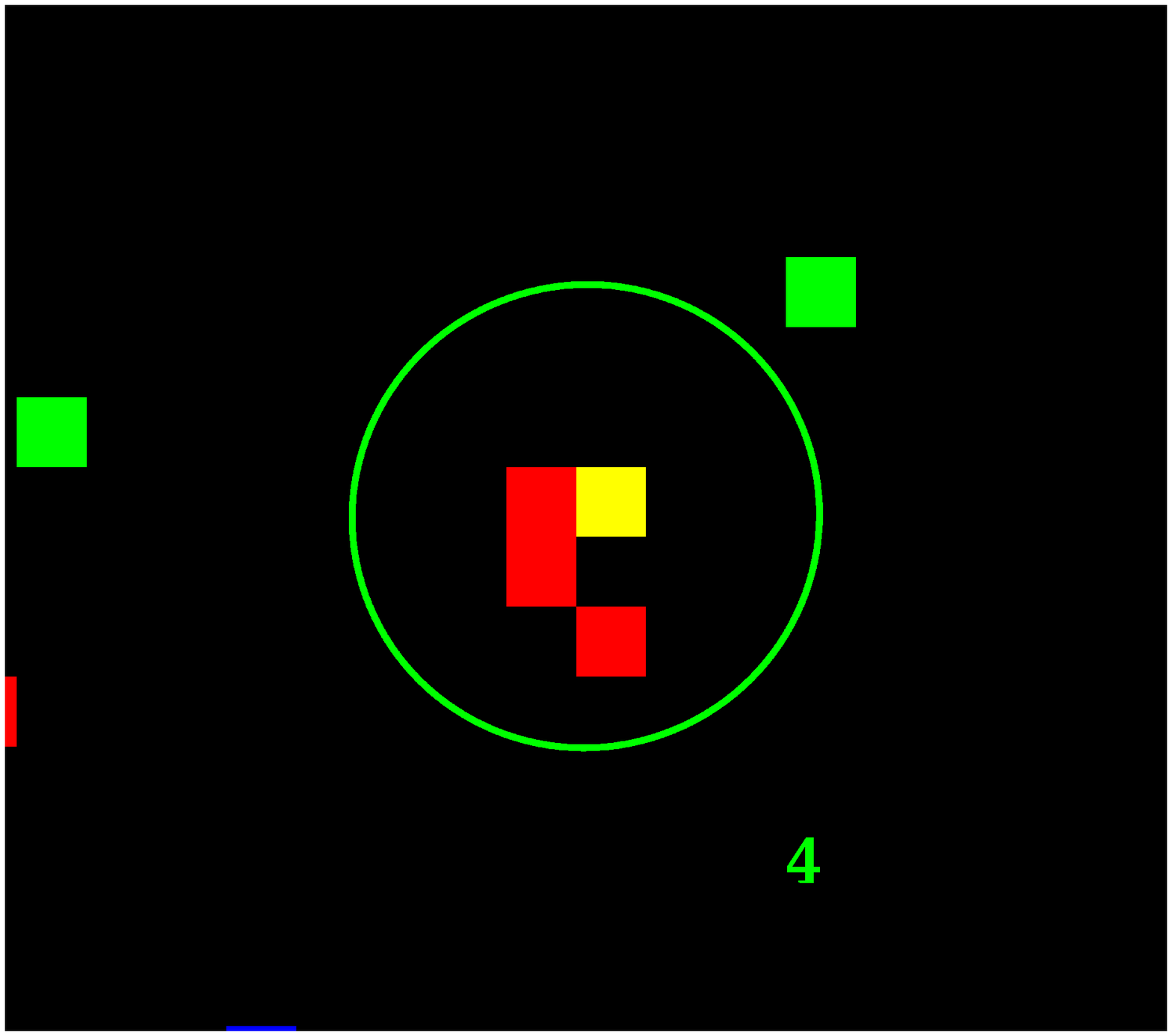}
\caption{{\em Chandra} postage stamps of 4 gas belt exclusion regions. Pixels are 0.492 arcsecond square, and data are displayed according to their energy (red 0.2-1.5\,keV, green 1.5-2.5\,keV, blue 2.5-8.0\,keV). The positions of the three regions are shown in Figure \ref{fig:dssgaszoom}.}
\label{fig:chandra_gas_ps}
\end{center}
\end{figure}

\begin{figure}
\begin{center}
\includegraphics[width=2.7in]{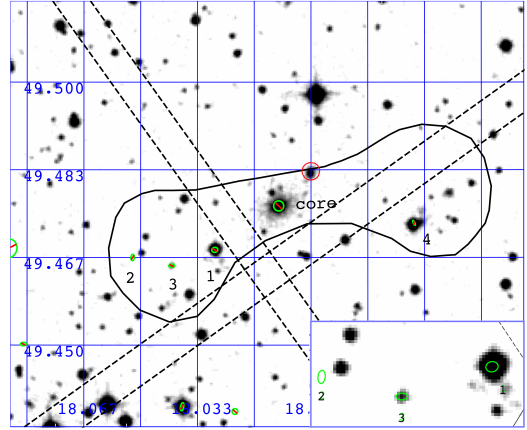}
\caption{{\em DSS} red image at 6100\AA (Second Palomar Observatory Sky Survey (POSS-II)) of the gas belt. Overlaid are positions of point sources detected with the {\sc ciao} task {\sc wavdetect}. The outline of the gas-belt region as defined by the X-ray data is shown in black and the {\em Chandra} chip gaps lie within the dashed parallel black lines. X-ray point source 3 in the gas belt is identified as a galaxy by eye. Gridlines are spaced as in Figure \ref{fig:xmmgaszoom}. An adjacent galaxy undetected in X-rays is marked by the red circle. {\em Inset}: Zoomed image of regions 1-3; 2. has no optical counterpart. }
\label{fig:dssgaszoom}
\end{center}
\end{figure}

In addition to the core, four point sources unresolved in the {\em Chandra} data (Figure \ref{fig:chandra_gas_ps}) and detected with {\sc wavdetect}, are embedded within the gas belt. We examined the Digitized Sky Survey ({\em DSS}, Figure \ref{fig:dssgaszoom}) and found that one of these X-ray sources is associated with extended optical emission (source 3). Sources 1 and 4 are confirmed as foreground stars, whereas 2 has no optical detection. 

The {\em Hubble Space Telescope (HST)} Wide Field Planetary Camera 2 (WFPC2)\footnote{http://www.stsci.edu/hst/wfpc2/} 5500\AA\,\, images of 3C\,35 are available in the {\em HST} archive (proposal ID 6967, P.I Sparks, W). We examine the calibrated (cosmic-ray free) science image, and confirmed the object marked within the red circle in Figure \ref{fig:dssgaszoom} to be a galaxy, probably a companion of the host galaxy of 3C\,35. This is further discussed in \S\ref{subsubsec:originofgasbelt}. 

\section{Spectroscopy}
\label{sec:spectroscopy_results}
X-ray spectra were fit with XSPEC 12.7.0 \citep{Arnaud96}, using $\chi^{2}$ statistics with Gaussian errors over the energy range 0.4-7.0\,keV. Metal abundance was fixed at $0.3Z_{\odot}$, and a Galactic neutral hydrogen column density of $1.23\times10^{21}$\,cm$^{-2}$ \citep[obtained with the \textit{Chandra} COLDEN\footnote{http://cxc.harvard.edu/toolkit/colden.jsp} tool, ][]{Dickey90} was adopted in all fits. Quoted errors on parameter values correspond to 90 per cent confidence for one interesting parameter ($\chi^{2}_{{\rm min}}+2.7$, all other interesting parameters are allowed to vary). The spectral index for non-thermal components, $\alpha$ is defined as $S_{\nu} \propto \nu^{-\alpha}$. The photon spectral index is $\Gamma = \alpha+1$, and the number power-law spectral index of radiating electrons is $p=2\alpha+1$ if the radiation has a synchrotron or inverse-Compton origin.
\subsection{Background Estimation}
\subsubsection{Chandra}
\label{subsubsec:chandra_back}
The radio structure of 3C\,35 (about $13'$ in the north-east to south-west direction) extends across all four chips in the \textit{Chandra} ACIS-I field of view, and the emission from the lobes is faint and diffuse, making the selection of an appropriate background region difficult. We examined spectra extracted using three distinct background regions; a small local background region close to the lobes (possibly including cluster gas), a larger local background, and a blank sky background event file from the \textit{Chandra} CALDB. The three analyses gave statistically consistent results, with the large local background producing the best fits and smaller parameter errors. Consequently, only the spectral results obtained with the large local background region (Figure \ref{fig:loberegions}) are described in this paper. 
\subsubsection{XMM}
\label{subsubsec:xmm_back}
We reduced the flare-free \textit{XMM} data using the XMM-ESAS software 
package. We chose to model the background rather than use a local 
background region due to the large extent of 3C\,35 and the spatial 
variability of the background; in particular instrumental emission lines are not 
uniform across the detectors. Blank-sky files were also inappropriate; residual soft proton and Solar Wind Charge eXchange (SWCX) contamination are highly variable \citep{Kuntz2008a}, and the periods of high flaring during our observation make calibration to the blank-sky files difficult. We modelled the off-source background using spectra extracted from a rectangular off-source region, west of the lobes of 3C\,35 and away from the gas-belt and group-gas extraction regions. Our background model was then included in fitting the source emission. 

\begin{figure}
\begin{center}
\includegraphics[width=3.20in]{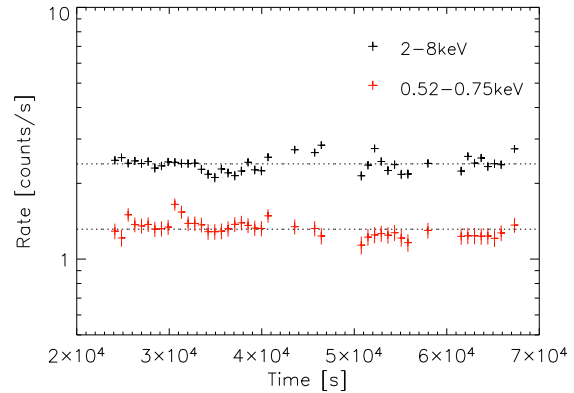}
\caption{\textit{XMM} pn light curve of the full field of view (after filtering for periods of soft proton flaring, see \S\ref{subsec:obsxmm}) for the 0.52-0.75\,keV band and the 2-8\,keV band in counts s$^{-1}$. Each light curve was created for events with data flag {\sc FLAG}=0, i.e events not from or near a bad pixel, and binned by 720\,s. The X-ray count rates do not have the particle background subtracted. The mean count rate per bin was corrected for any reduction in exposure of that bin, and the error shown on each data point is the standard error on the mean ($\sigma/\sqrt{N}$). The hard band shows some minor scatter, indicating a low level of residual contamination by soft proton flares. The soft band includes the majority of the [OVII] and [OVIII] line emission, which is associated with SWCX. At roughly $5\times10^{4}$\,s, there is a significant drop in the 0.52-0.75\,keV band, highlighting the variability of the SWCX emission.}
\label{fig:pnswcxa}
\end{center}
\end{figure}

We follow the prescription of \citet{Snowden2011} to model the \textit{XMM} background component. The XMM-ESAS task {\fontfamily{pcr}\selectfont mos\_back} generates a model quiescent particle background (QPB) spectrum using data from the FWC observations and unexposed corners of the MOS CCDs (similarly, {\fontfamily{pcr}\selectfont pn\_back} generates files for the pn CCDs). The quiescent instrumental background is excised from the data, while additional components must be modelled explicitly. Instrumental Al K$\alpha$ and Si K$\alpha$ lines were fitted with an unabsorbed Gaussian of zero intrinsic width at energies 1.49 and 1.75\,keV for the MOS detectors, and Al K$\alpha$ for the pn (the energy range for the pn data was restricted to $0.4<E<7.2$\,keV due to difficulties in modelling the QPB at high energies). SWCX comprises a significant fraction of the X-ray background at $E<1$\,keV, and it is usually observed in the form of a long-term enhancement during an observation. SWCX emission originates in the heliosphere; the high charge state ions in the solar wind interact with neutral atoms, thereby gaining an electron in a highly excited state. The electron subsequently decays by emission of X-rays, producing a background that is strongly dependent on the solar wind proton flux and heavy ion abundances \citep{Snowden2004, Koutroumpa2011}. Prominent SWCX emission lines include those from [O\,VII] and [O\,VIII] (see the ESAS cookbook and table 3 of \citet{Carter2011}). As suggested by \citet{Snowden2004}, we examined the light curve in the 0.52-0.75\,keV band and the 2-8\,keV band (Figure \ref{fig:pnswcxa}). The low energy curve (which contains the majority of the [O\,VII] and [O\,VIII] emission) showed a systematic drop in the average intensity after about $5\times10^{4}$\,s compared to the hard band, indicative of a change in state of the SWCX. We therefore include the emission lines for the ten lines listed in Table \ref{table:emissionlines} in subsequent modelling. 

\begin{table}
\begin{center}
\caption{ Principal emission lines used in the model, }
\begin{tabular}{@{}lclc}
\hline
\hline
Ion &	Energy (keV) &Ion &	Energy (keV)\\
\hline

C\,{\sc v} 	&0.299 & O\,{\sc viii} 	&0.653\\
C\,{\sc vi} 	&0.367 &Ne\,{\sc ix}	&$\approx0.9$\\
N\,{\sc vi} 	&0.420 &Ne\,{\sc x} 	&1.022 	\\
N\,{\sc vii} 	&0.500 &Mg\,{\sc xi} 	&1.330 \\
O\,{\sc vii} 	&0.561 &Si\,{\sc xiv} 	&2.000  \\
\hline
\end{tabular}
\label{table:emissionlines}
\end{center}
\end{table}

Diagonal response matrices and a power-law, not folded through the instrumental effective areas, were used to model the residual SP contamination (Figure \ref{fig:lobespecxmm}, the solid dark blue lines show the contribution of the fitted residual SP background for each detector). 
The cosmic X-ray background (CXB) was modelled by an unabsorbed thermal component representing emission from the heliosphere ($E\approx0.1$\,keV), an absorbed thermal component representing emission from the hotter halo/intergalactic medium ($E\approx0.25-0.7$\,keV) and an absorbed power-law of $\alpha\approx1.46$ to represent the unresolved background sources. We used the local RASS spectrum\footnote{The HEASARC X-Ray Background Tool can be found at http://heasarc.gsfc.nasa.gov/cgi-bin/Tools/xraybg/xraybg.pl} to constrain the contribution of the cosmic background (Figure \ref{fig:lobespecxmm}, lower blue line).

The best-fit model to the off-source data has $\chi^{2}= 98.4$ for 116 degrees of freedom. The best-fit indices and normalisations (and their 90 per cent errors) for the SP power-law and SWCX lines were used to constrain those background components for the on-source model, whilst the RASS data serve to constrain the contribution to the on-source fit from the CXB components. 
\subsection{The core}
\label{subsec:core}
We extracted the {\em Chandra} spectrum of the weak X-ray core from a circle of radius 1\arcsec (1.3\,kpc), using the same background regions as defined in \S\ref{subsubsec:chandra_back}. Although the statistics are poor (there are $9\pm3$ net counts, see Table \ref{table:regionparam}), we are able to derive a weak constraint on the nuclear luminosity. Assuming an unabsorbed power-law representing the nuclear X-ray continuum, modified by Galactic absorption \citep{Belsole2006}, the best-fit (using $C$-statistics) photon index and 1\,keV flux density are $\Gamma=1.2\pm1.0$, $S_{1\,\mathrm{keV}}=0.61^{+0.74}_{-0.39}$\,nJy. This corresponds to an unabsorbed 0.4-7.0\,keV luminosity of $8\times10^{40}$\, erg\,s$^{-1}$, which is much less than the average core luminosity of $L_{\mathrm{2-10\,keV}}>10^{44}$\,erg\,s$^{-1}$ found by \citet{Belsole2006} for 10 higher redshift FR\,II radio galaxies ($0.5<z<1.0$). The slope of the power law is consistent with the sample of 22 low redshift radio galaxies studied by \citet{evans2006}, who find that X-ray core components with low intrinsic absorption have a steeper mean photon index ($\Gamma=1.88\pm0.02$) than those with intrinsic absorption greater than $5\times10^{22}$\,cm$^{-2}$. The 5\,GHz flux density of the nucleus of 3C\,35 is roughly $12\,$mJy, based on our 1.4\,GHz map and a spectral index of 0.7. This places 3C\,35 in a typical $L_{\rm{X}}/L_{\rm{R}}$ location for the population of galaxies that have a nuclear component with intrinsic absorption less than $5\times10^{22}$\,cm$^{-2}$ \citep[][fig 2b]{evans2006}. We conclude that there is no evidence for a luminous hidden-core X-ray component in 3C\,35. 

Assuming all the core counts are associated with a cool corona and not emission from the AGN, we also fitted the core with a thermal model. The data are best fit with an {\sc apec} temperature of $3.1^{+64}_{-1.6}\,$keV. Although the upper error is unconstrained, there is no indication of a cool core. We froze the temperature to that found in the gas belt ($kT=1.28$\,keV, \S\ref{subsubsec:chandragas}), and found a 0.5-2.0\,keV luminosity of $2.5\times10^{40}$\, erg\,s$^{-1}$, although the fit was poor. This would place 3C\,35 at the extreme high-luminosity end of the 1.4\,GHz population of radio galaxies that host an X-ray corona \citep[][fig 6a]{sun2007}. The core is unresolved in the {\em Chandra} data and the counts are concentrated within $<1$\,kpc of the nucleus. Therefore, the core X-ray emission is most likely dominated by an active nucleus, and is unlikely to be associated with a cool core.
{
\renewcommand{\arraystretch}{1.4}
\begin{table*}
\begin{center}
\caption{Properties of the regions used for spectral extraction. Net count rates are measured in 0.4-7.0\,keV. For {\em Chandra}, the number in brackets is the fraction of counts in the extraction region attributed to the source. For the {\em XMM} data, `net' refers to the source, instrumental line, SWCX, SP, and CXB emission, minus the QPB background, i.e. only the QPB is excised from the data before fitting. Area is the net solid angle of the region, after excluding chip gaps, damaged CCDs and background sources. } 
\begin{tabular}{@{}c|cccc|cccc} \\ \hline
\hline
 		& \multicolumn{4}{|c|}{Net count rate ($10^{-2}$\,ct\,s$^{-1}$)}  & \multicolumn{4}{|c|}{Area (arcmin$^{2}$)} \\
		& {\em Chandra} & MOS1 & MOS2 & pn & {\em Chandra} & MOS1 & MOS2 & pn \\
\hline
Lobes		&  $2.75\pm0.23$ (23.8$\%$)	& $6.59\pm0.21$& $6.91\pm0.20$& $17.35\pm0.38$	& 31.3 & 24.4 & 23.7& 30.3\\
Gas belt	&  $0.50\pm0.11$ (17.3\%)	& $1.21\pm0.08$&$1.29\pm0.07$&$3.41\pm0.16$	& 3.58 &4.45& 4.38& 4.06 \\
Core		&  $0.035\pm0.017$ (98.5$\%$)	& - 	&-&-					& 0.001 &-&-&-\\
Group gas	&  $0.54\pm0.43$ (2.1$\%$)	& $1.63\pm0.1$&$1.46\pm0.1$&$4.15\pm0.21$	& 94.7& 12.6&12.6&12.6\\
\hline 		
\end{tabular}
\label{table:regionparam}
\end{center}
\end{table*}
}

\subsection{Lobes}
\subsubsection{Chandra}
\begin{figure}
\includegraphics[width=3.20in]{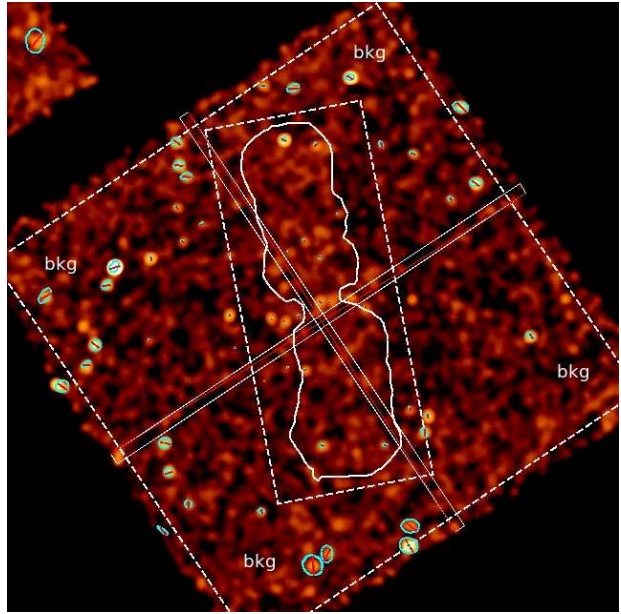}
\caption{\textit{Chandra} image of 3C\,35 in the energy range 0.4-7.0\,keV. Data are binned by 4 and smoothed with a Gaussian function of $\sigma=8$\,pixels ($\approx16\arcsec$). A radio contour of 0.5\,mJy/beam is used as the source region for spectral extraction. The background (labelled bkg) is defined as the region outside the large dashed rectangle centred on 3C\,35, but within the dashed square. Point sources and chip gaps are excluded from both regions. }
\label{fig:loberegions}
\end{figure}
\begin{figure}
\begin{center}
\includegraphics[width=3.0in,angle=270]{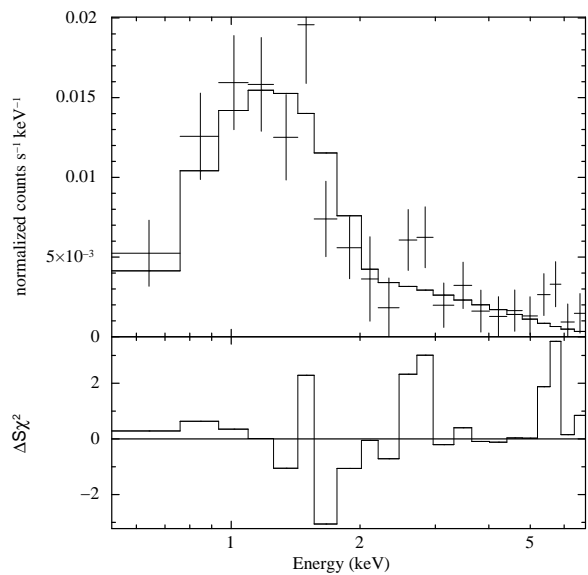}
\caption{\textit{Chandra} background subtracted, 0.4-7.0 keV spectrum of the diffuse emission fitted
with a power-law model plus Galactic absorption (Table \ref{table:lobeparam}). The lower panel shows the residuals' contribution to $\chi^{2}$. }
\label{fig:lobespecchandra}
\end{center}
\end{figure}

The \textit{Chandra} spectrum of the lobe X-ray emission was integrated within a radio contour of 0.5\,mJy/beam (delineated in Figure \ref{fig:loberegions}), excluding point sources and chip gaps. Figure \ref{fig:lobespecchandra} shows the spectrum of the lobes, after subtracting a background spectrum using the background region defined in Figure \ref{fig:loberegions}. Data are grouped to a minimum of 30 counts per bin. The net count rate and area of the extraction region are given in Table \ref{table:regionparam}. 

The lobes are best fitted with a power-law model, with parameters as given in Table \ref{table:lobeparam}. The best-fit photon index $\Gamma=1.79^{+0.39}_{-0.34}$ is consistent with that of the radio spectrum ($0.6<\alpha_{{\rm r}}<0.8$ between 73.8 and 327.4\,MHz, O10).
Under the assumption that $\alpha_{{\rm X}}=\alpha_{{\rm r}}$ (in common with e.g., IS11), we fix $\Gamma=1.7$ and add an {\sc xspec apec} model to the fit to model any thermal gas in the lobe region. We find a slightly improved fit with the addition of an {\sc apec} component ($\chi^{2}/$dof=20.15/19), although the uncertainties are such that the normalisation of the {\sc apec} model is consistent with being zero. 
{
\renewcommand{\arraystretch}{1.4}
\begin{table}
\begin{center}
\caption{Summary of the power-law model fitting to the lobes of 3C\,35 for {\em XMM} and {\em Chandra}. The Galactic neutral hydrogen column density is frozen for all fits. }
\begin{tabular}{@{}p{2.5cm}cc}
&&\\
\hline
\hline
 Model parameter 				& {\em Chandra} 			& {\em XMM}  \\
\hline
$N_{{\rm H}}$\,($10^{22}$\,cm$^{-2}$) 		& 0.123 			& 0.123 \\
$\Gamma$					& $1.79^{+0.39}_{-0.34}$	& $1.88^{+0.11}_{-0.11}[^{+0.17}_{-0.16}]$\\
$S_{\mathrm{1\,keV}}$ (nJy)			& $41.28^{+7.55}_{-7.28}$	& $47.78^{+2.92}_{-2.98}[^{+6.69}_{-6.29}]$	\\
$\chi^{2}/$dof					& 22.11/20 			& 214.75/213	\\
\hline 		
\end{tabular}
\label{table:lobeparam}
\end{center}
\end{table}
}
\subsubsection{XMM}
\label{subsubsec:xmmlobes}
\begin{figure}
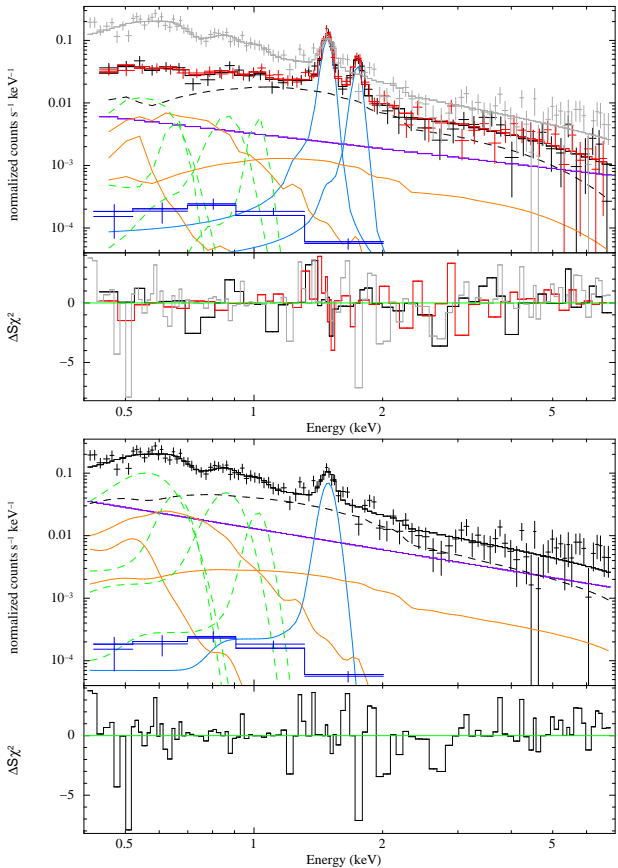

\begin{center}
\includegraphics[height=3.20in,angle=270]{3c35ic_combined_mos.ps}
\includegraphics[height=3.20in,angle=270]{3c35ic_combined_pn.ps}
\caption{{\em XMM} data and best-fit model of the diffuse emission from the lobes of 3C\,35 and X-ray background. {\em Top panel}: MOS on-source data from 0.4-7.0\,keV (pn in grey for comparison). The black and red lines are the composite MOS1 and MOS2 models respectively, the blue peaks are the Al\,K$\alpha$ and Si\,K$\alpha$ instrumental emission lines, the orange lines are the cosmic x-ray background components. The green dashed lines show the SWCX emission lines at [O\,VII] 0.565\,keV, [O\,VIII] 0.65\,keV, [Ne\,IX] 0.89\,keV, [Ne\,X] 1.053\,keV and the purple/dark blue solid line is the unconvolved power-law representing residual SP emission. The lower blue line and 5 blue crosses show the RASS data and best fit used to constrain the {\em XMM} CXB background. The dashed black line is a power-law representing the source emission from the region. The lower panel shows the residuals as their contribution to $\chi^{2}$. {\em Bottom panel}: pn off-source data from 0.4-7.0\,keV, model components as above.}
\label{fig:lobespecxmm}
\end{center}
\end{figure}

We extracted spectra integrated within a 0.5\,mJy radio contour of the lobes (excluding the contribution 
from point sources). The net count rates for each detector are given in Table \ref{table:regionparam}. We performed a combined fit of the EPIC MOS and pn spectra (Figure \ref{fig:lobespecxmm}), excluding 
energies outside $0.4<E<7.0$\,keV, and making use of the off-source fit to model the {\em XMM} background (\S \ref{subsubsec:xmm_back}). Table \ref{table:lobeparam} summarizes the parameters from the best fit. The first set of errors are statistical $1.65\sigma(90\%)$ values and the second set include $1.65\sigma$ errors on the background components also being fit. The source emission is well fitted by a power law with a photon index $\Gamma=1.88\pm{0.11}$ and a flux density at 1\,keV of $S_{\mathrm{1\,keV}}=48\pm3$\,nJy, which is consistent with the {\em Chandra} fit. 
Under the assumption that $\alpha_{{\rm X}}=\alpha_{{\rm r}}$, we fix $\Gamma=1.7$ and add an {\sc apec} model in order to test for thermal emission in the lobe region. Although hinted at in the {\em Chandra} data, the small component of thermal gas in the lobe region was not constrained. 
We froze the background components to the values for the power law alone, assumed their contribution to the error is zero, and derived a best-fit temperature of $1.6^{+2.0}_{-0.8}$\,keV and an absorption-corrected 0.4-7.0\,keV luminosity of $2.81^{+4.20}_{-2.79}\times10^{41}$\,erg\,s$^{-1}$; this indicates a thermal contribution to the total luminosity of the lobe region of roughly $7\%$ . This is consistent with the results for the gas belt in \S\ref{subsubsec:chandragas}. 

\citet{Floyd2008} use {\em H}-band ($1.6\,\mu$m) {\em HST} data to model the host galaxy of 3C\,35, and derive a best-fit half-light radius ($r_{{\rm e}}$) of 4.6\,kpc. If we assume emission from the host is truncated beyond $2r_{{\rm e}}$, the {\em XMM} PSF ($15\arcsec$) is large enough to include all X-ray emission from the volume of the optical galaxy. Since a region the size of the PSF was excluded from all spectral fitting, and our lobe fit is over the entire lobe region ($1$\,Mpc in diameter; large even for cluster gas) we interpret the detected thermal component as emission from the gas belt, which runs orthogonal to the lobe region and is visually the most prominent feature in the X-ray map after the lobes (see Figure \ref{fig:imgxmm}). 
\subsection{Gas belt}
\label{subsubsec:chandragas}
\begin{figure}
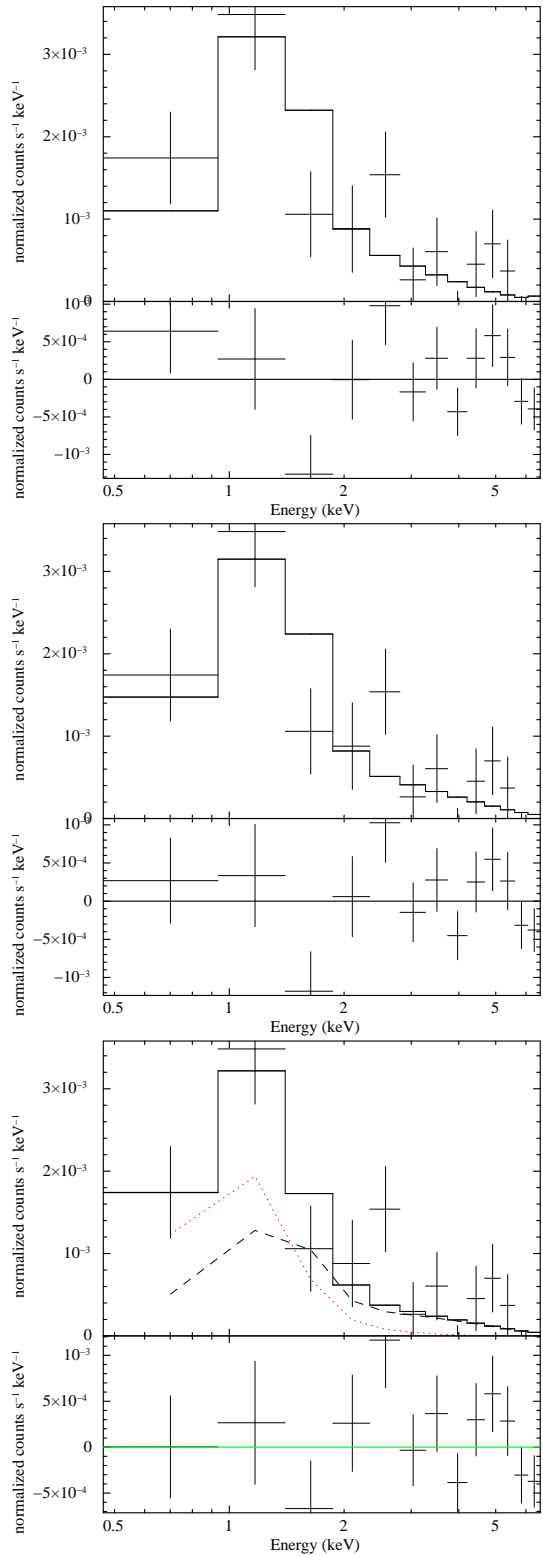

\begin{center}
\includegraphics[height=2.8in,angle=270]{chandra_belt_apec_only.ps}
\includegraphics[height=2.8in,angle=270]{gas_po_only_bw.ps}
\includegraphics[height=2.8in,angle=270]{apec_po_fix.ps}
\caption{{\em Chandra} background subtracted 0.4-7.0\,keV spectrum of the emission from the gas belt, fitted with {\em top} an {\sc apec} model, best-fit temperature of $\approx3.0\,$keV, {\em middle} a power-law with a free photon index, $\Gamma\approx2.2$ and {\em bottom} an {\sc apec} model (red dotted) plus power-law (black dashed) with a frozen $\Gamma=1.7$, best-fit temperature of gas $\approx1.0\,$keV. All models include Galactic absorption. (Table \ref{table:gasparam}). The lower panels show the fitting residuals i.e., the data minus the folded model.}
\label{fig:chandragasspec}
\end{center}
\end{figure}

We extracted {\em Chandra} spectra from a region we defined as the gas belt (Figures \ref{fig:imgxmm} and \ref{fig:xmmgaszoom}), orthogonal to the lobes of 3C\,35, excluding point sources and chip gaps. We have fitted several models to the 124 on-source, background-subtracted counts. Data are grouped into 32 bins of 32 channels width, and we used the $C$ statistic, which performs better than $\chi^{2}$ in the low-count regime \citep{Nousek1989}. 
{\em XMM} spectra were also extracted from the MOS1, MOS2 and pn detectors for the belt region. The net count rate for each detector is given in Table \ref{table:regionparam}. We modelled the {\em XMM} background as described in \S\ref{subsubsec:xmm_back}, making use of the off-source parameters to constrain the contribution of the background to the on-source fit. 
The {\em Chandra} spectra are shown in Figure \ref{fig:chandragasspec}, and best-fit parameters for both instruments are given in Table \ref{table:gasparam}. 

{
\renewcommand{\arraystretch}{1.4}
\begin{table}
\begin{center}
\caption{Best-fit parameters of the emission from the gas belt of 3C\,35 for {\em XMM} and {\em Chandra}. Models are absorbed {\sc apec}, power-law and {\sc apec} plus power-law. N is the {\sc apec} normalization quoted in units of $10^{-14}(1+z)^{2}\int n_{{\rm e}}n_{{\rm p}}dV/4\pi D^{2}_{L}$.}
\begin{tabular}{@{}p{3.cm}cc}
&&\\
\hline
\hline
 Model parameter 			& {\em Chandra} 			& {\em XMM}  \\
\hline
{\sc apec} 					&&\\
$kT$ (keV)				& $3.04^{+13.53}_{-1.85}$	& $2.47^{+0.50}_{-0.65}[^{+0.42}_{-1.73}]$\\
$N (10^{-5}$\,cm$^{-5}$)		& $4.79^{+1.22}_{-1.09}$	& $5.46^{+0.65}_{-0.65}[^{+0.99}_{-1.16}]$\\
$\chi^{2}/$dof				& cstat 			& 24.30/29	\\
\hline
{\sc power-law} 			&&\\
$\Gamma$				& $2.17^{+0.76}_{-0.65}$	&$2.11^{+0.22}_{-0.21}[^{+0.32}_{-0.20}]$		\\
$S_{\mathrm{1\,keV}}$ (nJy)		& $9.08^{+2.45}_{-2.65}$	&$10.20^{+1.19}_{-1.19}[^{+3.18}_{-2.65}]$	\\
$\chi^{2}/$dof				& cstat 			& 21.59/29	\\
\hline
{\sc apec + power-law} &&\\
$kT$ (keV)				& $1.04^{+1.36}_{-0.94}$	& $1.28^{+0.41}_{-0.30}[^{+1.59}_{-0.21}]$\\
$N_{\mathrm{apec}}$ ($10^{-5}$\,cm$^{-5}$)	& $1.36^{+2.04}_{-1.26}$	& $2.28^{+1.78}_{-1.20}[^{+1.66}_{-1.59}]$\\
$\Gamma$				& 1.7 (frozen)			& 1.7 (frozen)	\\
$S_{\mathrm{1\,keV}}$ (nJy)		& $5.04^{+3.58}_{-3.05}$	&$5.17^{+1.98}_{-2.38}[^{+2.25}_{-4.51}]$	\\
$\chi^{2}/$dof				& cstat 			& 19.97/28	\\
\hline
\end{tabular}
\label{table:gasparam}
\end{center}
\end{table}
}
The gas-belt {\em XMM} data fit a single {\sc apec} plasma model (with a fixed metal abundance at $0.3Z_{\odot}$) of temperature 2.5\,keV. We also fitted the data with an absorbed power-law, and find an acceptable fit with a photon index of $\Gamma=2.1$, suggesting the spectra cannot exclude non-thermal origin. A combination of an absorbed {\sc apec} + power law (with a fixed photon index $\Gamma=1.7$) is the best-fit model. This fit gives a lower temperature of the gas, $kT\approx1.0$\,keV, and smaller {\sc xspec} normalisation as compared to the single {\sc apec} model, and there is a good agreement between the {\em XMM} and {\em Chandra} measurements. The flux density of the power-law component is roughly $5.0$\,nJy, although the uncertainties are large. We attempted to allow for a varying $\Gamma$, but the fits then became unconstrained. 
\subsection{A detection of extended group gas}
\label{subsec:groupgas}
We extracted the {\em Chandra} spectrum from a large region surrounding 3C\,35 (within a circle of radius $400\arcsec$) but excluding all point sources in the field, the lobes and the gas-belt region; a net extraction region on the sky of approximately 95 square arcminutes. We chose the same background region as for the lobe spectral extraction, excluding the circular on-source region. The data were grouped into 32 bins of 32 channels width. We fitted an absorbed {\sc apec} model using the $C$-statistic, and found a best-fit temperature of $kT\approx1$\,keV, and an {\sc xspec} normalisation of $6.2\times10^{-5}$\,cm$^{-5}$ (corrected for excluded volumes). However, the group emission is weak, we could not constrain either of these values. Instead we assume these provide upper limits on the 0.4-7.0\,keV luminosity of the group gas encompassing 3C\,35, $L_{\mathrm{group}}\lesssim6.3\times10^{41}$erg\,s$^{-1}$ (corrected for excluded volumes). 

We also extracted the {\em XMM} spectra for a rectangular region ($12.6$ square arcmin, $4\arcmin$ to the west of the core) adjacent to the gas belt. We chose a small region offset from 3C\,35 rather than the 400\arcsec circle, due to the difficulty in modelling the {\em XMM} background over large areas. 
We modelled the background components in our extracted region as in \S\ref{subsubsec:xmmlobes} and derived best-fit parameter values assuming no error on the background components. The fit is good (reduced $\chi^{2}=0.77$). The {\em XMM} data are in good agreement with the {\em Chandra} data: we measure a group gas temperature of $kT=0.9^{+3.9}_{-0.4}$\,keV. We assumed the group gas occupies a sphere of radius of $400\arcsec$ centred on 3C\,35 ($V\approx1.4\times10^{67}$\,m$^{3}$) and assumed constant density, as the poor count rates prevent us extracting a density profile. We scaled the {\em XMM} measurement extracted from the smaller region up to this volume (a factor of roughly 10), accounting for projection effects in measuring the group-gas emission $4\arcmin$ away from the centre of the sphere. The 0.4-7.0\,keV luminosity of the group medium is then $L_{\mathrm{group}}=7.1^{+10.0}_{-3.3}\times10^{41}$\,erg\,s$^{-1}$. This value is consistent with {\em Chandra}, although the assumption of constant density used in scaling the {\em XMM} emission is very rough. 

{
\renewcommand{\arraystretch}{1.4}
\begin{table}
\begin{center}
\caption{Summary of the {\sc apec} fitting to the group gas surrounding 3C\,35 for {\em XMM} and {\em Chandra}. The Galactic neutral hydrogen column density is frozen for all fits. N is the {\sc apec} normalization quoted in units of $10^{-14}(1+z)^{2}\int n_{{\rm e}}n_{{\rm p}}dV/4\pi D^{2}_{L}$, corrected for excluded volumes. } 
\begin{tabular}{@{}p{4.0cm}cc}
&&\\
\hline
\hline
 Model parameter 				& {\em Chandra} & {\em XMM}  \\
\hline
$kT$ (keV)					& $\approx1$	& $0.9^{+3.9}_{-0.4}$\\
$N_{\mathrm{apec}}$ ($10^{-5}$\,cm$^{-5}$)	& $6.2^{+7.2}_{-6.2}$& $6.3^{+8.0}_{-3.9}$	\\
$\chi^{2}/$dof					& cstat		& 44.38/58	\\
\hline 		
\end{tabular}
\label{table:groupparam}
\end{center}
\end{table}
}

\section{Discussion}
\label{sec:discussion}
\subsection{Physical parameters of the lobes}
\label{subsec:lobediscussion}
The lobe X-ray spectra are well described by a power law (Table \ref{table:lobeparam}); the agreement between X-ray spectral index ($\alpha_{Chandra}=0.79^{+0.39}_{-0.34}$, $\alpha_{XMM}=0.88\pm{0.11}$) and the radio spectral index between 73.8 and 327.4\,MHz ($0.6<\alpha_{r}<0.8$) is consistent with the interpretation of the X-ray power-law component as inverse-Compton emission from a population of electrons in the lobes that is also radiating synchrotron radio emission. 
{
\renewcommand{\arraystretch}{1.4}
\begin{table}
\begin{center}
\caption{Physical parameters based on the X-ray emission from the lobes of 3C\,35. The photon index is fixed at the radio synchrotron index, errors correspond to 90 per cent confidence for one interesting parameter. $L_{\rm{X}}$ is the absorption-corrected $0.4-7.0$\,keV luminosity of the iC component. The errors on the derived physical parameters from the measurements are based on errors including the uncertainty on the background components. }
\begin{tabular}{@{}p{2.7cm}cc}
&&\\
\hline
\hline
 Parameter 				& {\em Chandra} 			& {\em XMM}  \\
\hline
$\Gamma$					& 1.7 (frozen)			& 1.7 (frozen)			 \\
$S_{\mathrm{1\,keV}}$ (nJy)			& $39.75\pm4.88$		& $42.41\pm2.25[^{+4.37}_{-4.50}]$	\\
$\chi^{2}/$dof					& 22.31/21 			& 214.39/216			\\
$L_{\mathrm{X}}$ ($10^{42}$\,erg\,s$^{-1}$)	&$3.51^{+0.43}_{-0.44}$		&$3.87\pm0.21[\pm0.36]$ \\
\hline 
$B_{\mathrm{me}}$ ($10^{-10}$\,T) 		& 1.58$\pm0.04$ 		& 1.58$\pm0.04$  \\
$P_{\mathrm{me}}$ ($10^{-14}$\,Pa)		& 0.72$\pm0.04$			& 0.72$\pm0.04$	\\
$B_{\mathrm{iC}}$ ($10^{-10}$\,T)		& 0.90$\pm0.12$			& 0.87$\pm0.1$	 \\
$u_{\mathrm{B}}$ ($10^{-15}$\,J\,m$^{-3}$)	& 3.22$^{+0.92}_{-0.80}$	& 3.01$^{+0.73}_{-0.65}$	 \\
$u_{\mathrm{p}}$ ($10^{-14}$\,J\,m$^{-3}$)	& 3.05$^{+1.12}_{-0.82}$	& 3.23$^{+1.03}_{-0.80}$ 	 \\
$P (10^{-14}$\,Pa)				& 1.12$^{+0.35}_{-0.24}$	& 1.18$^{+0.32}_{-0.24}$ 	\\
\hline
\end{tabular}
\label{table:lobephysics}
\end{center}
\end{table}
}

In order to diagnose energetics in the lobes, we fixed the radio synchrotron index at $\alpha_{{\rm r}}=0.7$. The number density spectrum of radiating electrons was assumed to be a power law of the form $\propto \gamma^{2\alpha_{{\rm r}}+1}$, where the Lorentz factor is $10^{2}<\gamma<10^{6}$. Source volume was calculated as $75\times10^{6}$\,kpc$^{3}$, and the radio flux density was estimated as $S_{\mathrm{R}}=2.4\pm0.2$\,Jy, based on our 1.4\,GHz VLA image. Referring to \citet{Worrall2006}, we derived a minimum-energy magnetic field $B_{{\rm me}}=0.16$\,nT (in agreement with O10) assuming a filling factor of unity, no relativistic protons, and no relativistic bulk motions. Table \ref{table:lobephysics} shows the minimum-energy parameters\footnote{At minimum energy, the energy in the magnetic field $u_{\mathrm{B}}=u_{\mathrm{p}}(\alpha+1)/2$, and total pressure (Pa) $P_{\mathrm{me}}=\frac{(3+\alpha)B^{2}_{\rm{me}}}{6(1+\alpha)\mu_{0}}$, where $B$ is in Tesla.}. 

We measured the corresponding inverse-Compton X-ray flux density (assuming CMB seed photons) from the power-law fit to the lobes with a photon index of 1.7. We assume all X-ray emission from this region is from the iC mechanism. The 1\,keV X-ray flux densities are shown in Table \ref{table:lobephysics}; we calculated the magnetic field in the lobes to be $B_{\mathrm{iC}}\approx0.09$\,nT, which is about 1.8 times lower than estimated under the minimum-energy condition, consistent with \citet{croston2005} who reported electron dominance ($B_{\mathrm{iC}}\sim0.7B_{\rm{me}}$) in the lobes of 33 other FR II radio galaxies and quasars. We then find the internal pressure in the lobes to be $1.2\times10^{-14}$\,Pa, slightly higher than predicted under the minimum-energy assumption. 

Using equations \ref{eq:n_p} and \ref{eq:P} (\S\ref{subsubsec:thermal+ic}), we calculated the confining pressure of the group environment as defined by the fits discussed in \S\ref{subsec:groupgas} to be $1.8^{+7.9}_{-1.0}\times10^{-14}$\,Pa. The lobes have comparable pressure with the external medium, within the uncertainties, although the assumption of hydrostatic equilibrium does not exclude some additional pressure source in the lobes. Strong electron or magnetic dominance are disfavoured because of the constraints from the iC emission. A relativistic proton component or decreased filling factor can provide additional pressure, as can extension of the electron spectrum to lower energies. A possible alternative source of pressure in the lobes is a population of old electrons near the inner regions of 3C\,35, steepening the electron spectrum. The power-law component of the gas-belt fit adds favour to this explanation, since we find it can be well described by increasing $\alpha$ towards the inner regions of 3C\,35. This is examined in more detail in \S\ref{subsubsec:thermal+ic}.

To make a comparison to the 33 radio sources analysed by \citet{croston2005}, we calculate $R$, the ratio of observed to predicted X-ray flux at equipartition. The 1.4\,GHz radio map predicts a minimum-energy magnetic field of $B_{\rm{me}}=0.16$\,nT and therefore $R\approx2.8$, i.e., the lobes are electron dominated. This falls within the narrow distribution of $R$-values found by \citet{croston2005}.

We measured the temperature of the group medium to be $kT=0.9^{+3.9}_{-0.4}$\,keV. The ratio of surface brightness, $\Sigma$, of the lobe X-ray emission to the surface brightness of the group medium differs by more than an order of magnitude, $\Sigma_{\rm{X}}^{\rm{lobes}}/\Sigma_{\rm{X}}^{\rm{group}}\approx30$. 
The group gas surrounding 3C\,35 has a bolometric luminosity of $L_{\rm{bol}}\approx10^{42}\,\rm{erg}\,\rm{s}^{-1}$ assuming a temperature of 0.9\,keV. \citet{croston2008} examined the luminosity-temperature relation for a sample of 9 low power FR\,I radio galaxies in group environments compared to radio-quiet groups. They found that groups containing extended radio sources have a higher temperature for a given X-ray luminosity. Their best fitting relation for the radio-quiet groups predicts a group gas temperature of $\approx0.6$\,keV for the group gas surrounding 3C\,35, below the measured value. Following \citet{croston2008}, we calculated whether it is possible for the radio source to raise the temperature of the group gas by 0.3\,keV, the difference between the observed group gas temperature and the temperature predicted by the radio-quiet $L_{\rm{X}}-T_{\rm{X}}$ relation of \citet{croston2005}. The energy required to raise the temperature of the group gas within a sphere of radius $400\arcsec$ encompassing 3C\,35 is E$_{\rm{req}}\approx4\times10^{52}$\,J. The total energy available from the current episode of radio activity, estimated as $4PV$ (this includes the energy stored in the lobes, which is yet to affect the environment) using the total lobe volume and the pressure in the lobes derived from the X-ray data (Table \ref{table:lobephysics}) is $10^{53}$\,J. Thus the radio source is sufficiently powerful to heat the group gas by $0.3\,$keV. However, as \citet{croston2008} point out, it is hard to exclude the possibility that, rather than the AGN driving the increase in the temperature of its environment, radio-loud AGN preferentially inhabit hotter than average group environments. 

\subsection{Comparison with Suzaku}
\label{subsec:suzaku}
{
\renewcommand{\arraystretch}{1.4}
\begin{table}
\begin{center}
\caption{A comparison of results from the {\em XMM} and {\em Chandra} data (assuming the best fit to the gas belt; {\sc apec} + power-law) with those from IS11 in the {\em Suzaku} fitting. IS11 attribute the soft thermal component of their fit to thermal plasma emission from the host galaxy of 3C\,35, which we compare to our measurements of the gas belt. }
\begin{tabular}{@{}p{2.9cm}ccc}
&&\\
\hline
\hline
 Parameter 					&  {\em XMM} 		&  {\em Chandra}  		& {\em Suzaku} \\
 						& 0.4-7.2\,keV		& 0.4-7.0\,keV			& 0.5-10.0\,keV		\\
\hline
Lobes + gas belt 				&&&\\
$L_{\rm{X}}$ ($10^{42}$\,erg\,s$^{-1}$)		& $4.1\pm0.4$		& $3.7\pm0.6$			& $2.2^{+0.8}_{-0.9}$			\\
Lobes 						&&&\\
$L_{\rm{X}}$ ($10^{42}$\,erg\,s$^{-1}$)		& $3.9\pm0.2$		& $3.5\pm0.4$			& $1.4^{+0.5}_{-0.6}$ \\
Gas belt/host galaxy 				&&&\\
$kT$ (keV)					& $1.3\pm0.4$		& $1.0^{+1.4}_{-1.0}$		& $1.3\pm0.3$			\\
$L_{\rm{X}}$ ($10^{41}$\,erg\,s$^{-1}$)		& $2.2^{+1.6}_{-1.1}$	& $1.8^{+1.7}_{-1.8}$		& $7.5^{+2.8}_{-3.7}$ \\
\hline
\end{tabular}
\label{table:isobecompare}
\end{center}
\end{table}
}
IS11 integrate the $0.5-10$\,keV \textit{Suzaku} spectrum within a large rectangular region containing the whole radio structure of 3C\,35. They report iC emission associated with the lobes, as well as a soft thermal component attributed to thermal plasma emission from the host galaxy. We have separated these two components based on higher resolution {\em Chandra} data and more sensitive {\em XMM} data. Table \ref{table:isobecompare} shows a comparison of the luminosities measured from each component across the three detectors. Both this paper and the IS11 analysis require careful subtraction of the X-ray background. Given this difficulty, the total X-ray luminosities measured for the lobes and gas belt of 3C\,35 are in reasonable agreement.

However, the 0.5-10.0\,keV absorption-corrected luminosity of the lobe iC component measured by IS11 is a factor of about 2 lower than the emission measured from the {\em XMM} and {\em Chandra} data. The IS11 thermal component is just consistent within errors with the {\em XMM} luminosity of the gas belt and we derive a gas-belt temperature similar to that of the IS11 thermal component.

In \S\ref{subsec:groupgas}, we reported an estimate of the luminosity of the group gas contained within a volume about 5 times larger than that of IS11; $L_{\mathrm{group}}=7.1^{+10.0}_{-3.3}\times10^{41}$\,erg\,s$^{-1}$ at a temperature of $0.9$\,keV. Therefore, there may be a contribution to the thermal component measured by IS11 from group gas, which would lower the luminosity attributed by IS11 to the host galaxy, bringing it more in line with the value we report for the gas belt. However, this does not explain the discrepancy between the iC measurements. We conclude that because of the poor spatial resolution of {\em Suzaku}, IS11 over-allocated power to the {\sc mekal} component of their fit and underestimated the total emission in the lobes, and included a contribution from group gas to the thermal component of their fit.

\subsection{The gas belt}
\label{subsubsec:thermal+ic}
It is clear from the fits in \S\ref{subsubsec:chandragas} that our spectra alone cannot prove the gas belt in 3C\,35 to have either a thermal or non-thermal origin. An {\sc apec} model with Galactic $N_{\rm{H}}$ describes the data (reduced $\chi^{2}=0.84$) and the derived parameters are consistent across the two instruments, although the temperature is not well constrained. Similarly a power-law model ($\Gamma=2.1$) also represents the data (reduced $\chi^{2}=0.88$). 
If we assume a thermal origin, the gas belt overlaps the lobe region by about $1.4$\,square arcminutes (Figure \ref{fig:xmmgaszoom}), and as such we might expect roughly 2\,nJy of the emission from the gas-belt region to be iC (assuming the flux density detected in the lobes). It was found that a power-law component added to the {\sc apec} model (with a spectral index consistent with the iC emission detected in the lobes) improved the fit.

The temperature of a purely thermal model ($kT\approx2.5$\,keV) is inconsistent with IS11, who find the temperature of the thermal component to be about $1.3$\,keV. Similarly, if the gas belt is of purely non-thermal origin, our results are inconsistent with IS11's measurement of a thermal component that is stronger than the group emission we see. In what follows, we discuss the physical plausibility of a gas belt emitting via a combination of mechanisms; {\sc apec}+power-law. Both the {\em Chandra} and {\em XMM} results are shown in Table \ref{table:gasparam}, but we use the more precise {\em XMM} results in subsequent analysis.

{
\renewcommand{\arraystretch}{1.4}
\begin{table}
\begin{center}
\caption{Temperature, density and pressure of the X-ray emission from the lobes, belt and group medium from the {\em XMM} data. The error in the volume is assumed to be 10\% for the lobes and extended group gas, and 50\% for the gas belt, to reflect our uncertainty on its shape. $L_{\mathrm{X}}$ is the absorption-corrected 0.4-7.0\,keV luminosity attributed to the thermal component of the gas-belt fit, the iC component of the lobe fit, and the thermal component of the group-gas fit, corrected for excluded volumes.}
\begin{tabular}{@{}p{2.7cm}ccc}
&&\\
\hline
\hline
 Parameter 				&  Lobes 			&  Gas belt  				& Group gas \\
\hline
$kT$ (keV)				& -			& $1.28^{+0.41}_{-0.30}$	& $0.9^{+3.9}_{-0.4}$		\\
Volume ($10^{66}$\,m$^{3}$)		& 2.2$\pm0.3$		& $9.3\pm4.6\times10^{-2}$	& 15.8$\pm1.6$			\\
$n_{{\rm p}}$ (m$^{-3}$)		& -			& $440\pm190$ 			& $60^{+35}_{-20} $		\\
$P (10^{-13}$\,Pa)			&0.12$\pm0.03$		& $2.0^{+1.1}_{-1.0}$		& $0.2^{+0.8}_{-0.1}$\\
$L_{X}$ ($10^{42}$\,erg\,s$^{-1}$)	&$3.9\pm0.2$		&$0.22^{+0.16}_{-0.11}$		& $0.7^{+1.0}_{-0.3}$		\\
\hline
\end{tabular}
\label{table:temppressure}
\end{center}
\end{table}
}
Making use of \citet{Worrall2006}, we calculate a density and pressure of the X-ray-emitting gas belt. The emission measure, defined in terms of the normalisation factor returned by {\sc xspec}, $N$, is 
\begin{equation}
10^{14}N=\frac{(1+z)^{2}\int n_{{\rm e}}n_{{\rm p}}dV}{4\pi D_{\mathrm{L}}^{2}}
\end{equation}
where $D_{\mathrm{L}}$ is the luminosity distance ($9.14\times10^{26}$\,cm), V is the volume (cm$^{3}$) and $n_{{\rm p}}$ (cm$^{-3}$) is the proton density. If $n_{{\rm e}}\approx1.18n_{{\rm p}}$ due to the presence of elements heavier than hydrogen in the gas, the proton number density for a region of uniform density is given by 
\begin{equation}
n_{{\rm p}}\sim\sqrt{\frac{10^{14}\mathrm{N}\,4\pi D_{\mathrm{L}}^{2}}{(1+z)^{2}1.18\,\mathrm{V}}}
\label{eq:n_p}
\end{equation}
if we assume that $n_{{\rm p}}$ is constant over the volume V. The pressure, $P$ (Pa) is given by
\begin{equation}
P = 3.6\times10^{-16}(n_{{\rm p}}\,kT)
\label{eq:P}
\end{equation}
where $n_{{\rm p}}$ is in m$^{3}$ and $kT$ in keV \citep{worrall2012}. We estimated the volume of the gas belt by assuming it is comprised of a disk of radius $r=100\,\arcsec$ and height $50\,\arcsec$. The physical parameters for the gas belt are in Table \ref{table:temppressure}.

\subsubsection{The origin of the gas belt}
\label{subsubsec:originofgasbelt}
The gas belt appears to be over-pressured compared to the group gas by an order of magnitude, although the errors are such that the belt and group gas could be in pressure balance. The belt is more clearly over-pressured (by an extra factor of about 2) if we do not allow for a power-law component. This suggests that the belt is expanded gas from the central regions, rather than the backflow of group gas displaced by the lobes. While these scenarios would suggest different temperature structures in the belt, the errors on the temperatures are not well constrained and so we cannot investigate the origin of the belt in this way. 

A comparison of the mass of gas in the belt with that expected in an elliptical galaxy provides a useful clue to its origin. An elliptical galaxy will typically contain a hot gaseous corona, ($kT\sim1$\,keV) which has been accumulated through mass loss during normal stellar evolution \citep{forman1985, matthews1990}. Galactic coronae are observed to be largely unaffected by their host AGN radio jets, so that conduction from the IGM into the coronae must be heavily suppressed \citep[][hereafter S07]{sun2007}. As \citet{osullivan2011} point out, heating efficiencies $\gtrsim1\%$ by AGN jets would raise the temperature of the X-ray coronal gas to that of the surrounding ambient medium, but the jets tunnel through the cool core with little interaction. Coronae, therefore, have posed problems for feedback models, since they can apparently provide enough gas to fuel an AGN for long periods, without being heated and swept away by the resultant radio jets. We investigate whether the belt may be a disrupted corona, the destruction of which would regulate the supply of distant gas to the central engine.

S07 studied the coronae of 157 early-type galaxies, 16 of which were identified with radio-loud AGN ($L_{\rm{1.4\,GHz}}>10^{23}$\,W\,Hz$^{-1}$). Of the 27 coronae that were resolved in the sample, S07 do not state what fraction host radio-loud AGN. For the resolved subset, they found a typical coronal radius of 1-4\,kpc, although several extend out to $\sim10\,$kpc. The coronal X-ray gas masses were generally in the range $10^{6.5}-10^{8}\,$M$_{\odot}$. Following their sample, we use the term `galactic corona' to refer to any component of gas with temperature $0.6\lesssim kT \lesssim1.6$\,keV, mass $\lesssim10^{8}$\,M$_{\odot}$ and a radius $\lesssim10$\,kpc.

We calculate the mass of the gas belt in 3C\,35 to be $M_{{\rm belt}}\approx(3\pm2)\times10^{10}$\,M$_{\odot}$, two orders of magnitude above the coronal masses in the S07 sample. One of the most massive resolved coronae in the S07 sample was that of NGC\,7720, with a gas mass of $\approx10^{9}$\,M$_{\odot}$. NGC\,7720 hosts a large FR\,I wide-angle tail source, about $400$\,kpc$^{2}$ in size, and has a high 1.4\,GHz radio power similar to that of 3C\,35. The discrepancy in mass estimates could be attributed to the larger size and higher X-ray luminosity of 3C\,35 compared to NGC\,7720, as well as extra mass accumulated via sweeping up IGM gas during expansion. 

\citet[][hereafter S09]{sun2009} expanded on the S07 sample of radio AGN hosting galactic coronae, presenting results for 52 galaxies with $L_{\rm{1.4\,GHz}}>10^{24}\,$W\,Hz$^{-1}$. Figure 1 of S09 shows that BCG (brightest cluster galaxies) in $kT<2\,$keV groups hosting galactic coronae fall in a region of $L_{\rm{X}}/L_{\rm{1.4\,GHz}}$ space separated from systems with large cool cores (LCC). The 0.5-2\,keV luminosity of the gas belt is $L_{\rm{belt}}=1.2^{+0.9}_{-0.6}\times10^{41}$\,erg\,s$^{-1}$, and the 1.4\,GHz flux of 3C\,35, $S=2.4$\,Jy, corresponds to a luminosity of $2.5\times10^{25}\,$W\,Hz$^{-1}$. If the gas belt originated from a corona, we would expect that during expansion its density, and therefore its luminosity, would have decreased. At earlier times, we would expect $L_{\rm{corona}}\appropto L_{\rm{belt}}V_{\rm{belt}}/V_{\rm{corona}}$ (the weak dependency on temperature is ignored). If we adopt the maximum radius of a corona measured in the S09 sample of coronae ($25\,$kpc)\footnote{although those systems with $r>10\,$kpc would probably be better classified as small cool cores}, then for 3C\,35 $L_{\rm{corona}}=(5\pm4)\times10^{42}$\,erg\,s$^{-1}$. S09 define a dividing line between systems with large cool cores (LCC, sources with cooling cores of cluster/group gas and a central isochoric cooling time of less than 2\,Gyr) and systems with galactic coronae, according to their 0.5-2.0\,keV luminosity; $L_{\rm{0.5-2.0\,keV}}\sim4\times10^{41}$\,erg\,s$^{-1}$. 3C\,35 is beyond the S09 boundary for galaxic coronae. While our calculated $L_{\rm{corona}}$ places 3C\,35 in the region allowed for a LCC, the extended group environment is weak (\S \ref{subsec:groupgas}) and the cooling time of the gas is much longer than 2\,Gyr (\S \ref{subsec:roleagnfeedback}), rendering a LCC implausible. 

We therefore conclude that the gas belt surrounding 3C\,35 is not of coronal origin, as it is over-luminous for a scaled back coronal radius of 25\,kpc and its mass is at least an order of magnitude larger than a typical corona of a single dominant elliptical galaxy. The optical host of 3C\,35 has an extensive envelope, and although luminous, is not atypical for a central elliptical galaxy. There is one other source outside the coronal class in fig. 1 of S09; 3C\,388, a FR\,II galaxy to be included in a future sample of `belted' radio sources (Mannering et al. in prep) exhibiting properties similar to 3C\,35.

An alternative interpretation of the gas belt is that it is accumulated gas lost from other galaxies in the group, or gas stripped from the host of 3C\,35 by surrounding group members. 3C\,285 ($z=0.079$) is another system with interesting structure in the thermal gas, also aligned orthogonal to the radio lobes \citep[][hereafter H07]{hardcastle07}. H07 refer to this structure as a `ridge', and measure a similar temperature of $kT=1.07^{+0.24}_{-0.11}$\,keV, a bolometric luminosity of $5.5\times10^{41}$\,erg\,s$^{-1}$ (twice that of the gas belt in 3C\,35), as well as a slightly smaller gas mass of $(2-3)\times10^{10}\,$M$_{\odot}$. H07 suggest the ridge is intrinsic to the system (rather than the interaction of the radio galaxy with the IGM), and may have even been present before the radio source switched on. It aligns well with the local distribution of galaxies and starlight (unaffected by the radio galaxy), so that a fraction of the gas in the ridge may have been stripped from the galaxies that merged to form the current host galaxy of 3C\,285. A case where an active merger aligns the gas content of several galaxies orthogonal to the radio lobes is 3C\,442A \citep{worrall2007}. 

To check the possibility that the gas belt in 3C\,35 is the combined ISM from several nearby galaxies we used the NASA Extragalactic Database (NED)\footnote{http://nedwww.ipac.caltech.edu} to search for galaxies within 6\arcmin\, ($450\,$kpc) of 3C\,35. A single galaxy is detected in the 2MASX survey\footnote{ 2 Micron All Sky Survey Extended objects, http://www.ipac.caltech.edu/2mass/}, at a separation of $5\arcmin$, and does not appear to be associated with 3C\,35 (although there is no redshift). By contrast, H07 find 12 objects classified as galaxies within a $5\arcmin$ ($450\,$kpc) radius of 3C\,285, 6 of which are optically confirmed (via SDSS) to be associated with membership of the group. 

We examined the galaxy\footnote{{\em RA}, $\delta$ = $01^{\rm{h}} 12^{\rm{m}} 00^{\rm{s}}.038$, $+49^{\rm{d}} 28^{\rm{m}} 58^{\rm{s}}.57$} previously identified in the {\em HST}/WFPC2 and {\em DSS} data (\S\ref{subsec:gasbelt_dss} and Figure \ref{fig:dssgaszoom}) using the Infrared Array Camera (IRAC) on board {\em Spitzer} \citep{fazio2004}. We find the new galaxy to be about $41\,$kpc offset from 3C\,35 (assuming it lies at a similar redshift), and at least 1.6 mag fainter. It is therefore likely to be a companion, but spectroscopic follow-up is needed to confirm this galaxy's association with 3C\,35. Thus, although 3C\,35 may be associated with a few fainter galaxies, the large magnitude difference suggests that it is relatively isolated, certainly compared to 3C\,285.

The potential companion galaxies of 3C\,35 are far too small to contribute significant gas or cause major disturbance in 3C\,35's gas. If the gas belt were a `ridge' in the IGM, we would expect to have detected an X-ray cool core in the host galaxy, due to the prevalence of such cores in radio-loud systems \citep{sun2007, sun2009}. We find no evidence for a cool core in the host of 3C\,35 (\S\ref{subsec:core}). We note that although H07 do not cite any evidence for a cool core in 3C\,285, there is evidence of a recent merger that may have disrupted a cool core.

A third possibility for the origin of the gas belt is that it is the residual of the expansion of fossil group gas, driven outwards by the radio lobes. Radio-loud AGN have been found to be prevalent in fossil group samples: \citet{hess2012} find 2/3 of their fossil group sample to be radio-loud. 3C\,35 roughly meets the standard criteria for a fossil group. Its closest companion is at least 1.6 mag fainter than 3C\,35 in the {\em Spitzer} IRAC data, and 3C\,35 has an extensive envelope which is not well fitted, so that the two galaxies may have a total brightness difference of about 2 mag. There are no other galaxies in the field of comparable luminosity. The combination of a luminous central galaxy \citep[the 2MASS $H$-band host galaxy magnitude of 3C\,35 of $-25.17$ \citep{skrutskie2006} is within the top few percent of the H-band luminosity function for local early-types,][]{bell2003} with a lack of neighbouring galaxies makes this system a convincing candidate for a fossil group. Fossil groups for which the temperature has been measured range from 0.66 to 4\,keV \citep{khosroshahi2007}, consistent with the temperature we measure for the gas belt.

\begin{figure*}
\begin{center}
\includegraphics[width=3.4in]{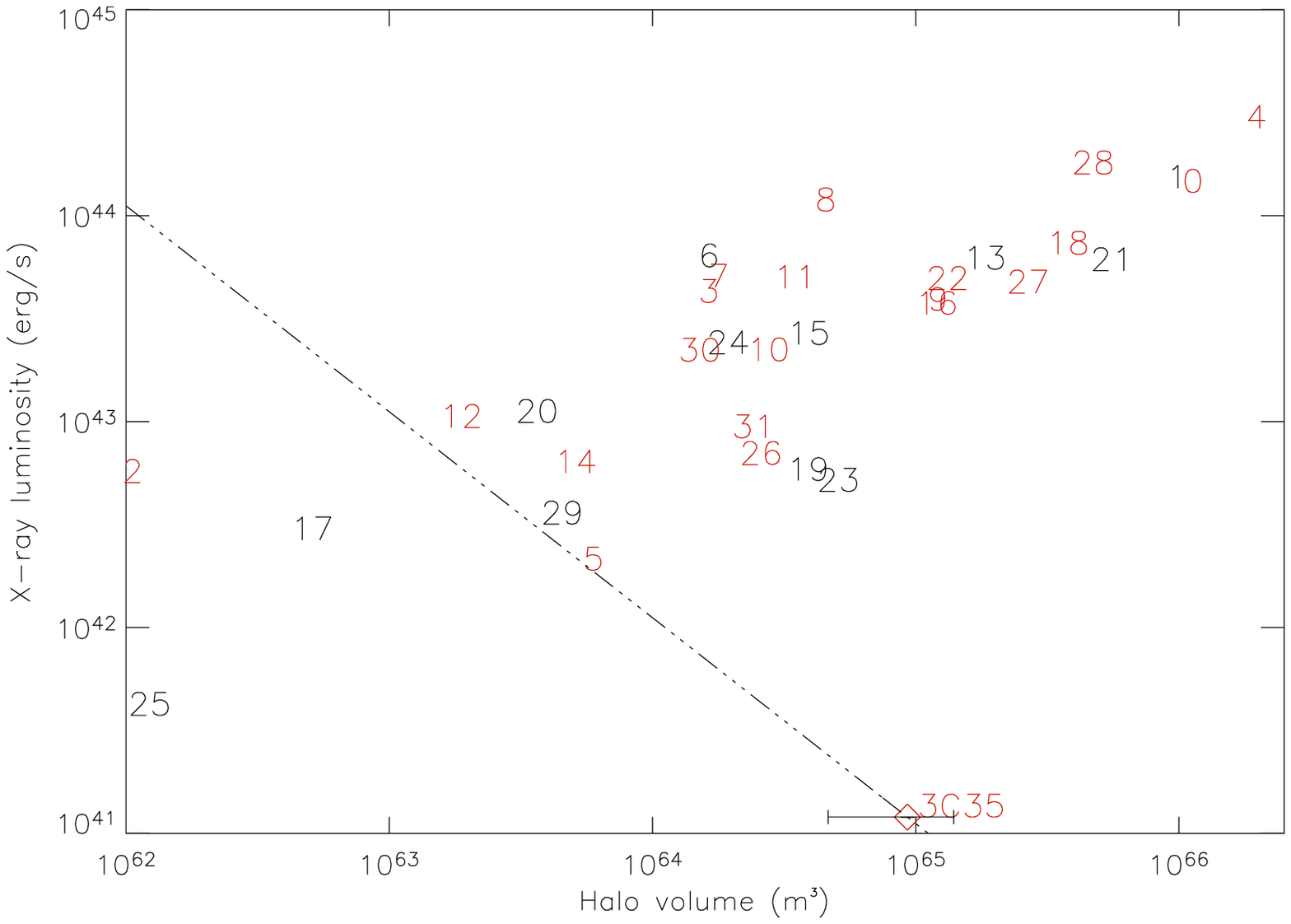}
\includegraphics[width=3.4in]{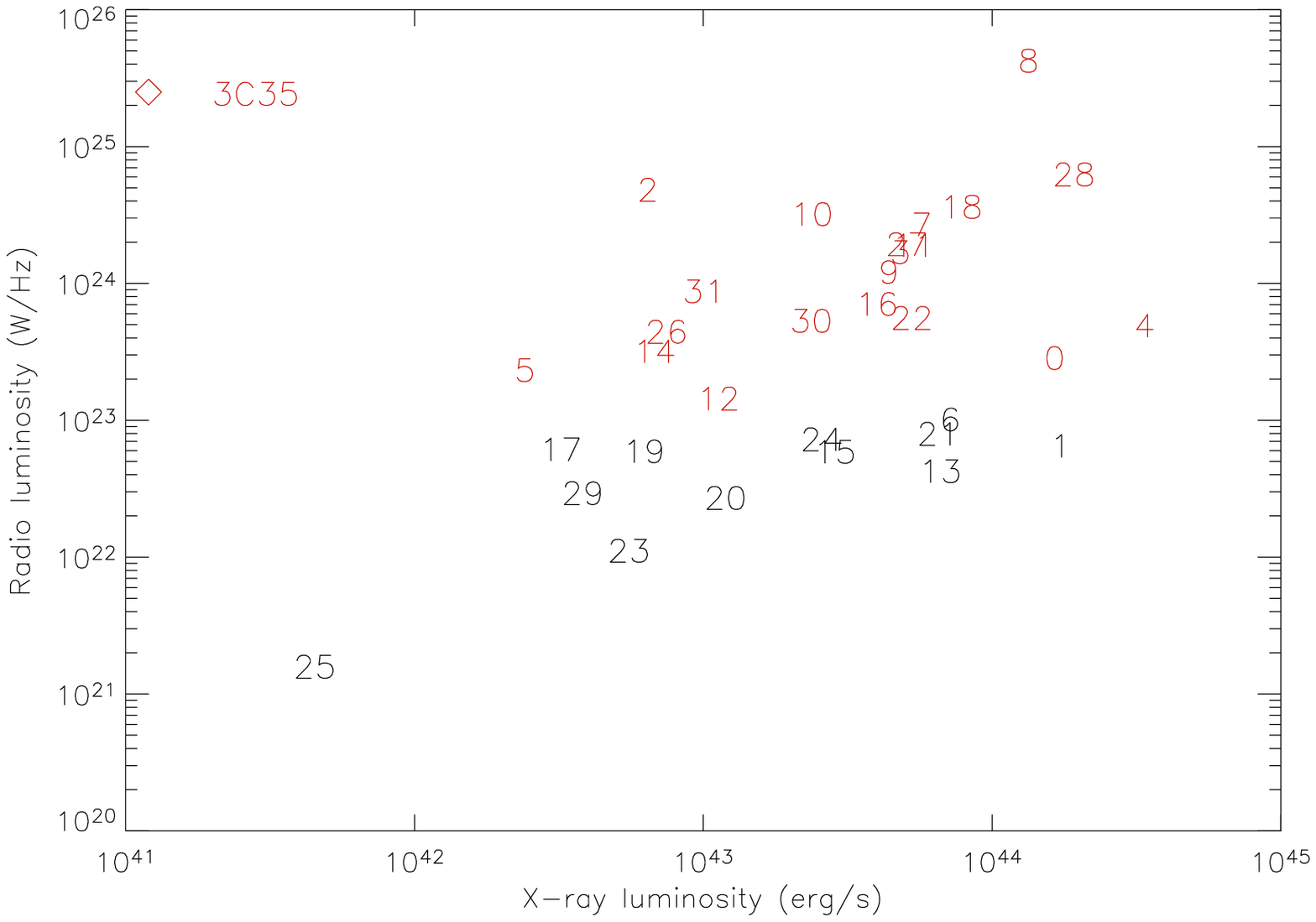}
\caption{{\em Left:} Halo 0.5-2.0\,keV luminosity against halo volume for fossil groups from \citet{hess2012, santos2007} with the gas belt of 3C\,35 marked with a diamond symbol. Those systems with a radio-loud ($>10^{23}\,$W\,Hz$^{-1}$) AGN are shown in red. The line shows the luminosity and volume of the gas belt at earlier times, assuming $L_{\rm{FG}}\propto L_{\rm{belt}}V_{\rm{belt}}/V_{\rm{FG}}$. The errors on the volume estimated for the gas belt reflect our uncertainty on its shape. {\em Right:} 1.4\,GHz radio luminosity against X-ray halo 0.5-2.0\,keV luminosity for the same objects.}
\label{fig:santosv_lx}
\end{center}
\end{figure*}

The 1.4\,GHz luminosity of 3C35 is $2.5\times10^{25}$\,W\,Hz$^{-1}$ (only one fossil group in the Hess sample is more radio luminous), yet the X-ray luminosity of the belt is low compared to the other fossil-group halos (Figure \ref{fig:santosv_lx}, right). But the belt volume for 3C\,35 is also large (Figure \ref{fig:santosv_lx}, left). If 3C\,35's belt arises from expansion of a fossil-group atmosphere, we would expect $L_{\rm{FG}}\appropto L_{\rm{belt}}V_{\rm{belt}}/V_{\rm{FG}}$, and this line is plotted in the Figure. The increased luminosity of the halo for plausible expansion factors of about 10 brings the system more in line with the Hess radio-loud fossil groups (Figure \ref{fig:santosv_lx}, right). 

Therefore, we believe the most likely explanation for the appearance of a relatively bright gas belt in such a poor environment is via the driven expansion of fossil group gas. We may be witnessing the extreme end of feedback, where the radio lobes have almost destroyed the atmosphere of a fossil group that fed the radio AGN.

The lobes and gas belt appear to be closely related morphologically (Fig. \ref{fig:imgxmm}), and have likely influenced one another during their formation; we expect comparable ages for both features. The pressure of the gas belt is higher than that of the lobes, thus the belt is capable of having driven the radio lobes out of the central regions of the source, explaining the ``pinch'' in the contours of the radio structure between the lobes (Fig. \ref{fig:lobes}). The pressure ratio (see Table \ref{table:temppressure}) between the gas belt and the group medium is $11^{+27}_{-10}$. Using the Rankine-Hugoniot conditions for a strong shock in a monatomic gas \citep[See \S5.1 of][]{Worrall2006}, this pressure ratio could drive supersonic expansion, although the implied Mach number, $3.0^{+2.3}_{-2.2}$ permits much slower expansions. 
The temperature and density predicted for the shocked group gas by expansion at a Mach number of 3, are allowed within the large errors\footnote{The predicted temperature ratio is about 3.7; we measure $T_{\rm{belt}}/T_{\rm{group}}=1.4^{+1.9}_{-1.2}$; similarly, $\rho_{2}/\rho_{1}\approx3.0$, and we measure $\rho_{\rm{belt}}/\rho_{\rm{group}}=7.8^{+9.2}_{-5.3}$}. 

The sound speed in the group gas around 3C\,35 is
\begin{equation}
c_{\rm{s}}=0.54\,(kT/\mathrm{keV})^{0.5}\quad\mathrm{ kpc\,Myr}^{-1}
\end{equation}
or $c_{\rm{s}} =0.5^{+1.1}_{-0.3}$\,kpc\,Myr$^{-1}$. Assuming a Mach number of 3, $v_{\rm{adv}}=1.6^{+3.4}_{-1.5}$\,kpc\,Myr$^{-1}$. The expansion timescale of the belt is then $t_{\rm{belt}}=80^{+180}_{-80}$\,Myr, for a belt radius of 130\,kpc (\S\ref{sec:morphology}). This is consistent with the synchrotron age of the radio source estimated by O10 of $t_{\rm{synch}}=143\pm20$\,Myr. The errors however, are too large to properly test the idea that the radio lobes and gas belt evolved simultaneously. 

A fit to the gas belt of purely thermal origin (Table \ref{table:gasparam}) finds a higher value of the temperature of $kT\approx2.5\,$keV, a proton density of $n_{{\rm p}}\approx680$\,m$^{-3}$ and a subsequent pressure of $P_{\mathrm{gas}}\approx6.0\times10^{-13}$\,Pa. The belt would then be over-pressured relative to the group medium by a factor of about 30. Following \S{\ref{subsubsec:thermal+ic}}, we derive a Mach number from the pressure ratio of gas belt to unshocked group gas of $\mathcal{M}\approx5$. We believe this to be unrealistic, as we find no evidence for shocks in the observations in the form of a bright edge to the gas belt.

\subsubsection{Fossil group or isolated elliptical?}
Some luminous elliptical galaxies are isolated, or reside in low-density environments, and may be mis-classified as fossil systems. \citet{jones2003} note that there are few field ellipticals with $L_{\rm{X}}>10^{42}$\,erg\,s$^{-1}$ that are not in the centres of groups or clusters and therefore have a contribution to their X-ray halo from IGM gas. \citet{mulchaey2010} study the $L_{\rm{X}}-L_{\rm{K}}$ relationship for a sample of field early-type galaxies, and detect an X-ray thermal component in all field galaxies with $K$-band luminosities above $L_{\rm{X}}\approx1.6\times10^{11}$\,L$_{{\rm K,} \odot}$ (more massive galaxies are able to retain their halos). The 2MASS $K$-band magnitude\footnote{over a $23.2\arcsec\times18.1\arcsec$ integration area} of 3C\,35 is 12.17 (corrected for Galactic extinction), which corresponds to $L_{\rm{K}}\approx2.45\times10^{11}$\,L$_{\rm{K,\odot}}$. Using the correlation of \citet{mulchaey2010}, a 0.5-2.0\,keV luminosity of $L_{\rm{X}}^{\rm{halo}}\approx(1.3-5.7)\times10^{40}\,$erg\,s$^{-1}$ is predicted for 3C\,35 from its stellar properties. We measure $L_{{\rm X}}^{{\rm belt}}=1.2^{+0.9}_{-0.6}\times10^{41}$\,erg\,s$^{-1}$.
A typical radius of the X-ray gas in elliptical galaxies is $30$\,kpc \citep{memola2009}, which corresponds to a volume 30 times less than that of the gas belt. If 3C\,35's belt arises by expansion of ISM in a field elliptical, then we would expect $L_{\rm{X}}^{\rm {halo}}=3.6^{+2.7}_{-1.8}\times 10^{42}$\,erg\,s$^{-1}$, an order of magnitude larger than the maximum $0.5-2.0\,$keV halo luminosity measured for field ellipticals \citep[$4\times 10^{41}$\,erg\,s$^{-1}$,][]{mulchaey2010}. Fossil groups have been shown to have systematically higher $L_{{\rm X}}$ and $L_{{\rm B}}$ than field ellipticals and non-fossil groups \citep{memola2009, hess2012}, thus we believe that the gas belt is likely to be accumulated ISM plus IGM from a merged group of galaxies, rather than a hot gas halo surrounding an isolated field elliptical. To firmly distinguish between the two possible evolutionary scenarios (fossil group or isolated galaxy), a detailed investigation of the dark matter potential around 3C\,35 is necessary.

\subsubsection{X-ray halo ejection and AGN fuelling}
The mass and age of the gas belt, as well as the morphological structure of the radio source provide a convincing argument that the central AGN activity has disturbed the gas morphology of the fossil group's X-ray halo, causing an ejection event and thus affecting the supply of large-scale gas to the central engine. We now address the physical plausibility of this scenario, given the power of the AGN outburst.

If we assume the $3\times10^{10}$\,M$_{\odot}$ of X-ray gas in the belt is being ejected from the galaxy, then it has been given a kinetic energy $U_{{\rm belt}}\approx10^{52}$\,J. The total energy available from the current episode of radio activity was estimated in \S\ref{subsec:lobediscussion} as $U_{{\rm lobes}}\approx10^{53}$\,J. Thus the lobes can easily supply enough energy to lift the belt gas. 

Following \citet{buttiglione2010}, we derived the black hole mass from the host luminosity using the correlation of \citet{marconi2003}
\begin{equation}
{\rm log}(M_{\rm{BH}})=-2.77-0.464\,M_{\rm{H}}
\end{equation}
where the optical host magnitude of 3C\,35 is $-25.17$ \citep[2MASS $H$-band,][]{skrutskie2006}, giving a black hole mass of $M_{\rm{BH}}=10^{8.9}$\,M$_{\odot}$. How much of the ejected gas must remain in order to fuel the AGN? To produce energies of $10^{53}$\,J in the radio structure, a mass of at least $10^{5.7}\,{\rm M}_{\odot}$ must have been accreted onto the central black hole. If the BH is accreting from a thin disk at 1$\%$ efficiency\footnote{An efficiency of up to about 0.4 is possible for $>10^{9}$\,M$_{\odot}$ BH \citep{davis2011}.}, $10^{7.7}\,{\rm M}_{\odot}$ is required after the ejection of the X-ray halo, i.e., less than $1\%$ of the gas mass of the belt, if left behind in the core of 3C\,35, would be sufficient to fuel the radio structure.

It is conceivable, that the dynamic gas in the belt originated as an X-ray halo of a fossil group, accumulated as a result of the merger of $L_{*}$ galaxies at $z\approx1$, and was driven outwards by the expanding radio jets before continuing to expand outwards at the sound speed. Its mass is comparable to that of the X-ray emitting IGM in a fossil group system, and the luminosities of the gas belt and radio galaxy are in line with brightest fossil group radio galaxies. The apparent influence of the radio structure on the belt implies the two have likely affected one another over the past roughly 150\,Myr. At earlier times the lobes would have been over-pressured relative to the belt, and probably shaped its formation. The similar evolution timescales of the radio source and gas belt support the idea that their evolution was coupled.

\subsubsection{The non-thermal component}
The non-thermal component required in fitting the gas belt implies a detection of inverse-Compton emission. The ratio of the surface brightness of the lobe X-ray emission (Table \ref{table:lobeparam}) to the X-ray surface brightness attributed to the power-law component in the belt is $\Sigma_{\mathrm{X}}^{\mathrm{lobes}}/\Sigma_{\mathrm{X}}^{belt}\approx1.0$. This is inconsistent with the radio data; from the VLA 1.4\,GHz radio map, the flux density in the gas-belt region is $30\pm2$\,mJy and so the ratio of lobe surface brightness to gas belt surface brightness $\Sigma_{\mathrm{R}}^{lobes}/\Sigma_{\mathrm{R}}^{belt}\approx9.8$. We would expect the X-ray and radio surface brightness ratios to be comparable, if the magnetic field strength and number densities of both the iC and synchrotron emitting electron populations are constant across the lobe and gas belt. In other words, in comparison with the lobes as a whole, the belt region appears to be more successful in producing power-law X-ray emission than would be expected on the basis of its radio flux. 

There are three possible scenarios to explain the variation of the X-ray/radio emission across the lobes and gas belt of 3C\,35. (i) another process is boosting X-ray emission from the central regions. (ii) a population of low-energy electrons in the central regions is contributing to the iC mechanism, or (iii) the magnetic field strength is varying as a function of position, such that the inner lobes are more electron-dominated.

Nuclear photons ($\gamma\sim10^2$) and starlight are important as extra sources of seed photons for iC scattering in some sources \citep[e.g.,][]{Brunetti97, stawarz2003}, and could provide a boost in the non-thermal X-ray emission seen from the belt of 3C\,35. 
Inverse-Compton scattering of nuclear photons should dominate at (projected) distances $R_{\mathrm{kpc}}$ from the core;

\begin{equation}
R_{\mathrm{kpc}}<70\,L_{46}(1-\mu)^{2}(1+z)^{-2}
\end{equation}

\noindent where $L_{46}$ is the isotropic nuclear luminosity ($10^{46}$\,erg\,s$^{-1}$) in the FIR to optical band and $\mu$ is the cosine of the angle between the radio axis and the line of sight \citep{Comastri03}. We assume $\mu=90\deg$ for 3C\,35. \citet{chiaberge2000} find an upper limit on the optical luminosity of 3C\,35 to be $L_{7000\,\mathrm{\AA}}<2.63\times10^{26}$\, erg\,s$^{-1}$\,Hz$^{-1}$ using Hubble Space Telescope (HST) data from the Wide Field and Planetary Camera 2 (WFPC2). 
Using infrared UKIRT observations of 3C\,35 \citep{lilly1985}, in four bands (JHKL) with fixed aperture size of 7.5\arcsec, we deduce the integrated IR to optical luminosity of the core region to be $L_{IR-opt}<1.85\times10^{44}$\,erg\,s$^{-1}$. 

Since the gas belt extends over 100\,kpc in radius, but $R_{\mathrm{kpc}}\lesssim1.3$, we conclude that iC scattering of nuclear photons does not contribute significantly to any power-law X-ray emission from the belt. It is also unlikely that a boost in gas-belt iC emission is due to the upscattering of starlight from faint galaxies in close proximity to 3C\,35, although we cannot rule out some iC emission originating from this process.

Scenario (ii), in which differences in the electron spectrum are responsible for the variation in X-ray/radio ratio, seems favourable. A population of lower energy electrons may exist in the gas-belt region. These would raise the X-ray iC emission, but not significantly contribute to the synchrotron mechanism at 1.4\,GHz. A steeper electron spectrum near the core (more older electrons) may add an additional component of pressure to the gas belt. A typical Lorentz factor for a population of electrons emitting synchrotron radiation at 1.4\,GHz in a fixed magnetic field of $0.09\,$nT is $\gamma\sim10^{4.4}$. In the case of CMB seed photons, the iC emission requires electrons with $\gamma\sim10^{3}$, the less energetic end of the synchrotron emitting population. 

Radio images of O10 (figure 1) support this explanation. The emission at low frequency shows a steeper radio spectral index near the core of 3C\,35: $\alpha_{r}$ changes from 0.6 to 1.7 between 327\,MHz and 1.4\,GHz in the inner region of the lobes. A steeper spectrum implies more low-energy electrons near the centre of 3C\,35 than we estimated in \S\ref{subsec:lobediscussion} above. If we increase the spectral index from $\alpha=0.7$ to $\alpha=1.2$, assuming the magnetic field strength measured in the lobes of $B_{\rm{iC}}=0.09$\,nT, we would expect to measure an increased flux at 1\,keV of 9.9\,nJy. This would give a surface brightness ratio of $\Sigma_{\mathrm{X}}^{\mathrm{lobes}}/\Sigma_{\mathrm{X}}^{belt}\approx0.6$.

Model (iii) postulates a variation in the magnetic field strength with position. A lower magnetic field could reduce the synchrotron emissivity in the belt, but the electron population could remain the same throughout the belt and lobe, accounting for the relatively uniform iC surface brightness. 
The observed X-ray and radio flux density are accounted for by iC scattering of the CMB in the belt if the magnetic field intensity is decreased to $B_{\mathrm{iC belt}}=0.02$\,nT (for $\alpha=0.7$), a factor of about 8 lower than predicted for minimum energy. This would imply that the magnetic field strengths are closer to equipartition in the more distant regions of the lobes as compared to the inner regions. 

It is unlikely that magnetic field variations between the lobes and gas belt of 3C\,35 or a population of lower energy electrons can independently explain the uniform X-ray flux but apparent decreased radio flux across the belt and lobes of 3C\,35. We speculate the most likely explanation is a combination of model (ii) and (iii), in which differences in both the electron spectrum and magnetic field strength in the belt are responsible for the variation in emission. Our fits can be explained if the electron spectrum is steepened to $\alpha=1.2$ and the magnetic field is increased from the lobe value to 0.12\,nT in the gas-belt region. This is only a factor of $30\%$ higher than the magnetic field measured in the lobes, and about a factor of two less than predicted by a minimum-energy calculation. 

\subsection{The role of AGN feedback in fossil groups}
\label{subsec:roleagnfeedback}
In this paper, we have suggested that 3C\,35 is destroying the surrounding IGM. We have found that the gas belt has an age comparable to that of the radio source and that the morphologies of the radio lobes and gas-belt point to co-evolution. The gas belt is unlikely to be asymmetrically aligned group gas accumulated from galaxies in the plane of the belt, as 3C\,35 resides in a poor environment, neither is it consistent with disrupted galactic coronal gas. We favour the argument that the gas belt is expanded fossil group gas, driven outwards by the radio lobes. We have demonstrated that the radio source is powerful enough to unbind the group gas from the central galaxy potential well, and that only 1\% of the mass of the belt is required to have fuelled the radio structure.

We find the pressure of the external group medium to be up to two orders of magnitude lower than the gas belt, implying the belt is an expansion of gas from the central regions, rather than a backflow of heated IGM displaced by the radio lobes. However, without robust temperature measurements across the gas belt, we cannot completely rule out the latter scenario. A backflow could provide fuel to the AGN, continuing its outburst. This would be the type of feedback loop on scales of hundreds of kpc necessary for radio sources to respond to their large scale environments \citep[a weakness of feedback models has been the absence of a responsive mechanism over large distances,][]{dubois2010}. 

The detection of radio-loud AGN in fossil groups \citep[][and this work]{hess2012} implies these systems are not as quietly and passively evolving as previously thought. The short lifetime of radio synchrotron emission (less than a few hundred\,Myr) after the source has turned off \citep[][and references therein]{sun2004} compared to the fossil group formation timescale inferred from simulations (several Gyr), as well as the long cooling time for the X-ray emitting gas \citep[$t_{\rm{cool}}$ of the gas belt is estimated as $20-50$\,Gyr;][figure 5]{Worrall2006} suggests that a reccuring process drives AGN activity.

The mechanisms governing the fuelling of AGN, and the processes involved in turning nuclear activity on and off in these dynamically old systems, remain unclear. In 3C\,35 it appears not to be merger-driven. We see no stucture in the $HST$ and $Spitzer$ images that might support a major merger in this system for more than about 1 Gyr. The optical nuclear spectrum of 3C\,35 \citep{buttiglione2009} is absorption-line dominated, with no sign of optical AGN activity or a recent strong burst of star formation (in the last $\approx 20$\,Myr).

A relationship between AGN and their host galaxies has been well documented \citep[BH-bulge mass correlation, e.g.][]{ferrarese2000} and discussed in terms of AGN-induced feedback through several proposed mechanisms; AGN-driven winds, starburst-driven superwinds, and jet-induced outflows \citep[e.g.,][and references therein]{holt2008,hambrick2011}. AGN feedback remains the most likely mechanism to balance radiative cooling of the IGM in systems with large cool cores \citep[][and references therein]{sun2012}. There has been no clear example of feedback between radio sources hosted in fossil systems and the large scale gas in the literature. 3C\,35 could be the first case where we are seeing the X-ray halo of a fossil group during the end phase of disruption. 

\section{Conclusion}
We have discussed the extended X-ray emission seen in new {\em XMM} and archival {\em Chandra} observations of the FR II giant radio galaxy 3C\,35. The properties of the lobes are consistent with earlier work (IS11, O10), and show a clear detection of inverse-Compton emission. The implied departure from equipartition is comparable to the range observed in other sources. We report detections of an extended group-scale environment, and place weak constraints on the non-thermal emission from the nucleus. 

More importantly, we report the detection of a gas belt, orthogonal to the radio lobes and lying between the radio lobes seen in our 1.4\,GHz radio map. We conclude that the X-ray emission from the belt is most likely a combination of thermal and power-law emission. The thermal component is likely to be from hot gas originating in $\sim L_{*}$ galaxies that merged over a Gyr ago and was driven outwards by the expanding radio structure before continuing to expand at its own sound speed. The age of the radio structure of 3C\,35 is consistent with the lifetime of the expanded belt feature. The higher than predicted (from 1.4\,GHz data) X-ray flux of its non-thermal component is attributed to a population of low energy electrons boosting iC emission in combination with positional variation of the magnetic field. 3C\,35's power is sufficient to place it close to the peak of the distribution of radio power in the local universe, and thus in the critical range where radio-mode feedback should be seen \citep{worrall2009}, if it is indeed an important mechanism in the required regulation of structure evolution. 

In order to support the mechanism being common and to determine a possible age/power/environmental dependency on the ejection of the X-ray halos of fossil groups by radio-loud AGN requires a sample of potentially similar cases. We are currently investigating a sample of nine such sources (Mannering et al. in prep). Simulations would provide beneficial insight into this phenomenon.

\section{ACKNOWLEDGEMENTS}
This paper is based on observations obtained
with \textit{XMM-Newton} and \textit{Chandra}.
\textit{XMM-Newton} is an ESA science mission with instruments
and contributions directly funded by ESA Member States and
NASA. We thank the CXC for its support of Chandra observations and 
data analysis, and the SAO R\&D group for DS9 and
FUNTOOLS. 
This work has also used data from the VLA.
NRAO is a facility of the National Science Foundation operated
under cooperative agreement by Associated Universities, Inc. EM thanks the STFC for support.

\end{document}